\documentclass[%
reprint,
onecolumn,
superscriptaddress,
amsmath,amssymb,
aip,
]{revtex4-1}

\usepackage{graphicx}% Include figure files
\usepackage{dcolumn}% Align table columns on decimal point
\usepackage{bm}% bold math
\usepackage[utf8]{inputenc}
\usepackage{mathptmx}
\usepackage{etoolbox}
\usepackage{subfigure}
\usepackage[T1]{fontenc}
\usepackage{hhline}
\usepackage{multirow}
\usepackage{float}
\usepackage{makecell}
\usepackage{comment}
\usepackage{mathtools}
\usepackage{xcolor}
\usepackage{mathrsfs}  
\newcommand{\blue}{\color{black}} %blue!70!black
\newcommand{\black}{\color{black}}

\begin{document}

%\preprint{AIP/123-QED}
%\begin{CJK*}{UTF8}{gbsn}
\title{Log-law recovery through reinforcement-learning wall model for large-eddy simulation} %[LL recovery through RLWM for LES]

\author{Aurélien Vadrot}
\affiliation{Department of Mechanical and Production Engineering, Aarhus University, 8200 Aarhus N, Denmark }
\author{Xiang I.A. Yang}
%\author{Xiang I.A. Yang ({杨翔})}
\affiliation{Department of Mechanical Engineering, Pennsylvania State University, State College, PA, 16802, USA}
\author{H. Jane Bae}
\affiliation{Graduate Aerospace Laboratories, California Institute of Technology, Pasadena, CA, 91125, USA}
\author{Mahdi Abkar}
\email{abkar@mpe.au.dk}
\affiliation{Department of Mechanical and Production Engineering, Aarhus University, 8200 Aarhus N, Denmark }

\date{\today}

\begin{abstract}

This paper focuses on the use of reinforcement learning (RL) as a machine-learning (ML) modeling tool for near-wall turbulence. 
RL has demonstrated its effectiveness in solving high-dimensional problems, especially in domains such as games. 
Despite its potential, RL is still not widely used for turbulence modeling and is primarily used for flow control and optimization purposes.
A new RL wall model (WM) called VYBA23 is developed in this work, which uses agents dispersed in the flow near the wall.
The model is trained on a single Reynolds number ($Re_\tau = 10^4$) and does not rely on high-fidelity data, as the back-propagation process is based on a reward rather than output error. 
The states of the RLWM, which are the representation of the environment by the agents, are normalized to remove dependence on the Reynolds number. 
The model is tested and compared to another RLWM (BK22) and to an equilibrium wall model, in a half-channel flow at eleven different Reynolds numbers ($Re_\tau \in [180;10^{10}]$). 
The effects of varying agents' parameters such as actions range, time-step, and spacing are also studied. 
The results are promising, showing little effect on the average flow field but some effect on wall-shear stress fluctuations and velocity fluctuations. 
This work offers positive prospects for developing RLWMs that can recover physical laws, and for extending this type of ML models to more complex flows in the future.

\end{abstract}

\maketitle
%\end{CJK*}

%\linenumbers

\section{Introduction}

Machine learning (ML) techniques have been widely used recently in lots of various domains, including robotics \cite{kober2013reinforcement}, cybersecurity \cite{nguyen2021deep}, biology \cite{jumper2021highly}, games \cite{silver2018general} and others. 
Computational fluid dynamics (CFD) has not been spared. 
ML has been applied for turbulence modeling \cite{yang2019predictive,maulik2019sub,eidi2022data}, flow control \cite{ren2020active,guastoni2023deep}, reduced-order modeling \cite{hasegawa2020machine}, optimization \cite{viquerat2021direct,mekki2021genetic,rincon2023validating}, model discovery\cite{callaham2021learning,bakarji2022dimensionally}, among others. 
An overview of ML applications in CFD can be found in Refs. \citenum{duraisamy2019turbulence,brunton2020machine,vinuesa2022enhancing,zehtabiyan2022data}. 

ML elicits as much attention as it does criticism in the turbulence community.
First, ML is a subfield of artificial intelligence (AI) that focuses on the development of algorithms and statistical models that enable computers to learn from and make predictions or decisions based on data, without being explicitly programmed to do so.
ML encompasses a broad spectrum of techniques, ranging from the straightforward, such as linear regression, to the highly complex, such as reinforcement learning (RL). 
Within the context of turbulence modeling, the term "machine learning" often involves regression, for which the most commonly used tool is the feed-forward neural network (FNN).
This type of network processes information in a unidirectional manner, flowing from input nodes through hidden nodes to output nodes. 

ML is often seen as a highly effective and versatile solution, and it is applied to a wide range of problems.
In the context of modeling, ML is frequently perceived as a method for discovering the ideal model from data, without the need for prior hypotheses, at the expense of physical consideration \cite{legaard2021constructing}.
ML is frequently referred to as a "black box" because it is difficult to understand the connection between the inputs and the predictions produced by the model. 
This makes it challenging to form a simplified explanation, or hypothesis, about the relationships in the data. 
In contrast, traditional modeling techniques involve formulating hypotheses based on observations in a given theoretical framework, constructing a model that represents that hypotheses, and testing the model's ability to predict phenomena in this framework \cite{durbin2018some}.
ML often bypass data observations to produce a model, resulting in a lack of prior hypotheses and the discovery of multiple models that perform well on the training set but have unknown theoretical frameworks. 
When applied to new data, these models often perform poorly and produce unrealistic results \cite{bae2022scientific}, and we have limited control over the prediction. 
The solution in ML turbulence modeling involves incorporating physical knowledge to inform, constrain, and/or embed the model \cite{wu2018physics,zhang2019recent}.
ML is commonly perceived as a model in and of itself, but it is actually just a tool. 
A model should be based on true hypotheses within a physical framework. 
This does not mean that ML cannot be successful in turbulence modeling, as it is still in its early stages and many possibilities have not been explored yet. 
Despite its limitations, such as its black-box nature and limited extrapolation capabilities, ML might have the potential to overcome these challenges through the use of various types of neural networks \cite{kurz2023deep,deng2019super,lapeyre2019training,duvall2021discretization}. 
It is important to note that the above-mentioned observations and criticisms about ML are not applicable to all ML techniques and are more commonly made about ML in general rather than specific ML methods.

This paper aims to use a specific ML tool, reinforcement learning (RL), which has limited usage in turbulence modeling \cite{bae2022scientific,novati2021automating,kurz2023deep,kim2022deep} but has various applications in CFD for control and optimization \cite{guastoni2023deep,verma2018efficient}.
Novati \emph{et al.} \cite{novati2021automating} were the first to use RL for predicting sub-grid scale (SGS) turbulence model in the context of large-eddy simulations (LES) followed by Refs. \citenum{kim2022deep,kurz2023deep}. 
Xiang \emph{et al.} \cite{xiang2021neuroevolution} applied RL and accelerated convergence of numerical solutions of the Pressure Poisson equation with density discontinuities.
Later, Bae \& Koumoutsakos \cite{bae2022scientific} resorted to RL for predicting wall-shear stress ($\tau_w$) in the context of LES wall modeling. 

RL is an ensemble of algorithms to solve optimization problems. 
RL does not necessarily require high-fidelity data since the training is done on the fly with a trial-and-error approach. Working directly with \textit{a posteriori} data greatly help to generalize since the discrepancy between \textit{a priori} and \textit{a posteriori} data (caused by e.g. discretization and filtering errors) is a main source of errors \cite{duraisamy2021perspectives}. 
The basic idea here is to adjust an agent’s or agents’ behaviors in an environment to yield desired outcomes.

In the context of LES WM, the WM is the agent, the LES field is the environment, and the desired outcome is an
accurate wall-shear stress. 
The agent is able to learn a relevant policy to maximize its long-term reward. 
The reward can be local or global, coming from experiment, theory or high-resolution data. 
In RL, the reward, serving as the loss function, is not determined by comparing predicted and true outputs, unlike in supervised learning.   

RL seems promising in answering the two aforementioned concerns: the "black-box" nature and the limited extrapolation capabilities. 
Ref. \citenum{vadrot2022survey} gave promising clue to understand agents' behavior and "predict the prediction" by looking at the states-action map.
Moreover, the involvement of physical consideration into RL turbulence modeling is possible through the definition of states, rewards and actions.
The states are the only information about the environment that agents can see.
By constraining the states, we can embed physical knowledge in the representation of the environment built by agents.
Agents do not need to know all the flow field; and in fact by doing so, we construct a too-specific model that will not be able to adapt to other flows.
Bae \& Koumoutsakos \cite{bae2022scientific} compare the performance of two models: the velocity-based WM and the log-law-based WM. 
The latter, with states defined based on the intercept and slope of the log law, achieved better generalization. 
However, it is not yet perfect, a log-layer mismatch is shown in Ref. \citenum{vadrot2022survey} at small and large Reynolds numbers that does not result from directly using velocity at the first off-wall grid point \cite{yang2017log,kawai2012wall}.

The following paper aims to pursue this work in the context of WM. 
Modeling near-wall turbulence is a key element for LES. 
WM consists in a strong simplification of the flow behavior close to the wall. 
The richness of turbulence composed of streaks and quasi-streamwise vortices is reduced to a single value of the wall-shear stress ($\tau_w$).
The outer-layer flow is used to predict the wall-shear stress (or heat flux), which will be itself used to predict the outer-layer flow, following an iterative process. 
The most basic, yet widely used, form of WM – equilibrium wall model (EWM) \cite{schumann1975subgrid,kawai2012wall,yang2018semi,chen2022wall,xu2021assessing} – is based on a simplified solution of fluid flow near the wall, in which all the non-equilibrium terms (related to e.g., pressure gradient, acceleration, and buoyancy) are neglected.
While the EWM is generally effective in predicting the flow behavior, including in non-equilibrium flows \cite{larsson2016large}, its ability to accurately capture strong non-equilibrium effects in certain flow scenarios can be limited \cite{hansen2023pod,fowler2022lagrangian}.
Recent developments in wall modeling, e.g., the integral WM \cite{yang2015integral}, the non-equilibrium WM \cite{park2014improved}, the slip-WM \cite{bose2014dynamic,bae2019dynamic} and the Lagrangian relaxation towards equilibrium WM \cite{fowler2022lagrangian}, seek improvements by incorporating a lower level of simplifications. 
Further efforts are still needed to better predict complex flows with separation, transition, and heat transfer.
ML has paved the way for a new branch of WM research that promises to improve precision and complexity beyond what traditional wall modeling techniques can achieve.
The ever-increasing availability of high-fidelity simulation data \cite{perlman2007data,graham2016web,lee2015direct} have motivated the use of machine-learning wall models (MLWMs). 
The past few years have seen the development of a number of ML WMs \cite{yang2019predictive,huang2019wall,huang2021bayesian,zhou2021wall,zhouArxiv,bae2022scientific,zhou2022wall,lozano2020self,bhaskaran2021science,radhakrishnan2021data,moriya2021inserting}.
Yang \emph{et al.} \cite{yang2019predictive} was the first to apply ML in WM using supervised MLWM trained at $Re_\tau=1000$ to predict the wall-shear stress. 
Huang \emph{et al.} \cite{huang2019wall,huang2021bayesian} built upon Yang \emph{et al.}\cite{yang2019predictive}'s WM to develop WMs that work well in spanwise rotating channel and channel with arbitrary (in terms of direction) but small (in terms of magnitude) system rotation, while still recovering the law of the wall in basic channels \cite{vadrot2022survey}. 
Bin \emph{et al.} \cite{bin2022progressive} employed a progressive learning approach that emulates the development of empirical WM, gradually increasing the complexity of the flow.
Our approach follows a similar progressive methodology, now using RL.
Our primary objective is to precisely capture the log law in equilibrium flows prior to attempting more complicated flows.
We think it is crucial to proceed with care and not apply RL directly to complex flows without progressively recovering physical laws.

The paper presents an improvement to the original reinforcement learning wall model (RLWM) \cite{bae2022scientific}.
The new model, named VYBA23, is trained at a moderately high Reynolds number ($Re_\tau = 10^4$) and uses newly identified states to overcome the limitations of the original BK22 WM \cite{bae2022scientific}.
This new model demonstrates a successful recovery of the law of the wall, up to an extremely high Reynolds number ($Re_\tau=10^{10}$).
Moreover, some aspects of RL, such as the impact of the distance between agents, the time-step between actions or the range of actions, are still largely under-explored. 
This paper further investigates these effects.
The rest of the paper is organized as follows. The computational setup and RLWMs are detailed in Section \ref{sec:Detail}. The results are presented in Section \ref{sec:result} followed by an analysis in Section \ref{sec:analysis} and concluding remarks afterwards.

\section{Wall-modeled large eddy simulation details}
\label{sec:Detail}
\subsection{Flow configuration and flow solver numerics}

The configuration is the half-channel flow.
The domain size is $L_x\times L_y\times L_z = 2\pi \delta \times 2\pi \delta \times 1\delta$ in the streamwise $x$, spanwise $y$, and wall-normal $z$ directions, where $\delta$ is the half-channel height.
The flow is periodic in both the streamwise and the spanwise directions.
A wall-shear stress boundary condition is imposed at $z=0$ (wall), and a symmetric condition is imposed at $z=\delta$ (at the half channel height).
The flow is driven by a constant pressure gradient in the $x$ direction. 
The friction Reynolds number is varied in the range: $Re_\tau \in \left[180, 10^3, 2\times10^3, 5.2\times 10^3, 10^4, 10^5, 10^6, 10^7, 10^8, 10^9, 10^{10}\right]$.
Note that the RLWM has been trained using only one Reynolds number within this range ($Re_\tau=10^4$), and then tested for all Reynolds numbers.

We employ the open-source pseudo-spectral code LESGO, publicly available at \url{https://lesgo.me.jhu.edu} \cite{lesgo}.
The code uses the spectral method in the $x$ and $y$ directions and the second-order finite difference method in the $z$ direction.
The computational domain is divided uniformly into $N_x=48$, $N_y=48$, and $N_z=48$ grid points with the resolution of $dx$, $dy$, and $dz$ in the $x$, $y$, and $z$ directions, respectively.
The grid planes are staggered in the vertical direction, with the first horizontal velocity plane at a distance $dz/2$ from the surface. 
The last grid point is just above the physical domain, and therefore $N_z$ grid points translate to a wall-normal grid spacing of $L_z/(N_z-1)$.
The code has been well validated and extensively used in earlier research publications \cite{porte2000scale,abkar2012new,abkar2015influence,yang2018hierarchical2,yang2020scaling,yang2022logarithmic}.
Furthermore, it has served as a ground for testing SGS models and WMs \cite{bou2005scale,stoll2006dynamic,yang2015integral,abkar2016minimum,abkar2017large}. %moeng1984large,
Available SGS models include the constant coefficient \cite{smagorinsky1963general}, dynamic \cite{germano1991dynamic}, and Lagrangian dynamic \cite{meneveau1996lagrangian} Smagorinsky models, and the minimum dissipation model (AMD) \cite{rozema2015minimum,abkar2016minimum}.
Available WMs include the EWM \cite{bou2005scale,yang2017log}, the integral WM \cite{yang2015integral}, the slip-WM \cite{yang2016physics}, the POD-inspired WM \cite{hansen2022pod}, the supervised MLWMs in Refs. \citenum{huang2019wall,zhou2021wall} and the RLWM in Ref. \citenum{bae2022scientific}.

\subsection{\label{subsec:WM} Wall models}

Three WMs are considered, namely the equilibrium wall model, referred to as EWM \cite{moeng1984large}, the RLWM in Ref. \citenum{bae2022scientific}, referred to as BK22 and the newly developed RLWM in this paper, referred as VYBA23. The first two models underwent a thorough comparative analysis in Ref. \citenum{vadrot2022survey}. The latter will be thoroughly investigated and compared to the BK22 WM.

\subsubsection{Empirical WM, EWM}
\label{subsubsec:EWM}

The EWM imposes the following law of the wall locally and instantaneously:

\begin{equation}
\label{eq:log_law}
    u^+ = \frac{1}{\kappa} \ln \left(\frac{z}{z_0}\right),
\end{equation}
where $u^+=u/u_\tau$ is the inner scaled streamwise velocity, $\kappa \approx 0.4$ is the von Kármán constant, $z_0=\nu \exp(-\kappa B)/u_\tau$ is a viscous scale, and $B\approx 5$ is the intercept of the log law \cite{marusic2013logarithmic}.
The model reads:

\begin{equation} \label{eq:wall-shear}
    \tau_w=\rho u_\tau^2=\rho\left[\frac{\kappa \tilde{U}_{\rm LES}}{\ln(h_{wm}/z_0)}\right]^2,
\end{equation}
where $\rho$ is the fluid density, $U_{\rm LES}$ is the LES horizontal velocity at a distance $h_{wm}$ from the wall, and $ \tilde{\left( \cdot \right)}$ denotes possible filtration operation \cite{yang2017log}.
Equation \ref{eq:wall-shear} is implicit and must be solved iteratively.
The matching height $h_{wm}$ can be the first, second, or the n\textsuperscript{th} off-wall grid point \cite{kawai2012wall,zhou2021wall}. 
In this work, we place $h_{wm}$ at $dz/2$, i.e., the first off-wall grid point, and filter the LES velocity to remove the log-layer mismatch \cite{yang2017log}.
Here, the factor $1/2$ is due to the use of a staggered grid.

\subsubsection{Reinforcement learning wall model, BK22}

\begin{figure}
\centering
\includegraphics[width=0.8\textwidth]{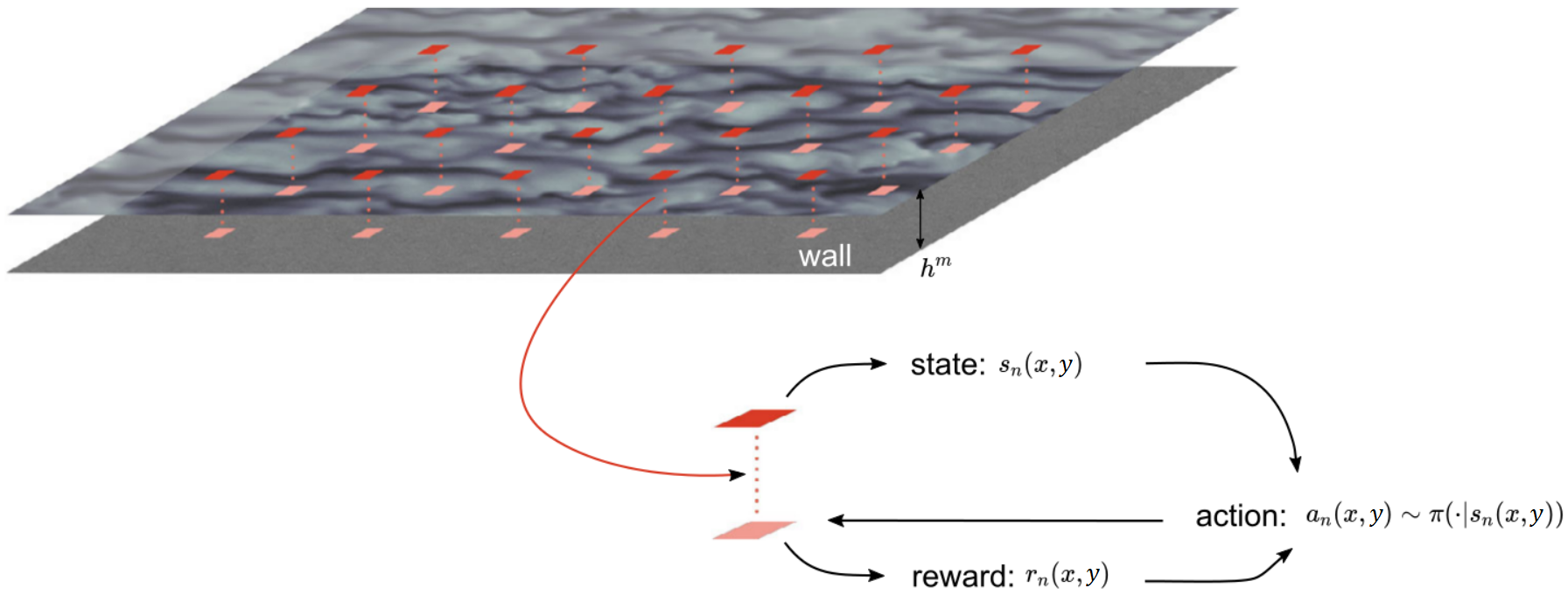}
\caption{\label{fig:schema_BWM} Several single-policy agents are distributed in the ($x,y$) plane at $h_{wm}$. Reprinted from Ref. \citenum{bae2022scientific}, licensed under a Creative Commons Attribution 4.0 International License.}
\end{figure}

Figure \ref{fig:schema_BWM} shows how the RLWM works.  
Bae \& Koumoutsakos \cite{bae2022scientific} initialized their training with EWM-generated flow fields at $Re_\tau=2000$, 4200, and 8000. 
The RLWM was trained using only the initial EWM-generated fields as initial condition, after which it acted independently without further guidance from the EWM-generated fields. 
Several agents are inserted into the LES flow field, and these agents modify flow fields in order to reach the best policy $\pi$.
The optimal policy is determined using a neural network featuring two hidden layers, each with 128 neurons, which generate the policy's mean value and standard deviation.
The network parameters have been optimized using an off-policy actor-critic algorithm known as V-Racer, with additional information available in Ref. \citenum{bae2022scientific}. 
Specifically, the agent produces an action $a_n(x,y)$ on its environment at the instant $t_n$ based on an observation (the states $s_n$) and a reward $r_n$, causing the environment to transition from states $s_{n}$ to states $s_{n+1}$. 
The action is to increase or decrease the predicted wall-shear stress by a factor of $a_n$ as:

\begin{equation}
    \tau_w(x,y,t_{n+1}) = a_n(x,y) \tau_w(x,y,t_n).
\end{equation}
In Ref. \citenum{bae2022scientific}, $a_n(x,y) \in [0.9, 1.1]$. We will test the effect of this range on the predictions in Section \ref{sec:result}.

Each agent receives a reward computed as:

\begin{equation}
    r_n(x,y,t_n) = \frac{|\tau^{\textrm{true}}_w-\tau_w(x,y,t_n)| - |\tau^{\textrm{true}}_w-\tau_w(x,y,t_{n-1})|}{\tau^{\textrm{true}}_w} + \textrm{Bonus}.
\end{equation}

$\tau^{\textrm{true}}_w$ is the true mean wall-shear stress obtained from the equations of motion for an unidirectional mean flow. 
The approach of converging to the theoretical mean value is a common wall modeling strategy in LES, such as the equilibrium wall model, which enforces the law of the wall locally and instantaneously.
Since LES involves unsteady and dynamic processes, connecting the wall-shear stress with a fluctuating LES velocity can result in varying wall-shear stress, even when the law of the wall is imposed. 
The objective of RLWM is to accomplish the same effect.
In a channel flow, the mean wall-shear stress at the lower wall ($z=0$) is related to the mean streamwise pressure gradient as: 

\begin{equation} \label{eq:true_tauw}
    \tau^{\textrm{true}}_w = -\delta \frac{dp_0}{dx}.
\end{equation}

In LESGO, the flow is driven by a constant pressure gradients,  $\tau^{\textrm{true}}_w$ is thus known.
The bonus reward helps to accelerate the convergence of the model if expressed as:

\begin{equation}
\textrm{Bonus} =  \begin{dcases*} 
 1-\frac{|\tau^{\textrm{true}}_w-\tau_w(x,y,t_n)|}{\tau^{\textrm{true}}_w},  & if  $|\tau^{\textrm{true}}_w-\tau_w(x,z,t_n)|/\tau^{\textrm{true}}_w < 0.1$; \\ 
  0 & otherwise. 
  \end{dcases*} 
\end{equation}

It is worth noting that the agents in Figure \ref{fig:schema_BWM} are single-policy agents.
That means that from given states, all agents will predict the same action. 

In Ref. \citenum{bae2022scientific}, two models are trained: one called the velocity wall model (VWM), and one called the log-law wall model (LLWM). 
We consider only LLWM, which was proven to have better extrapolation capabilities because of its definition based on the empirical log law. 
The states ($s_1$ and $s_2$) of LLWM are respectively defined as:

\begin{subequations}
    \begin{equation}\label{eq:kappa}
         s_1 = \frac{1}{\kappa_{wm}}=\frac{h_{wm}}{u_{\tau_{wm}}} \left.\frac{\partial u}{\partial z}\right|_{z=h_{wm}}, 
    \end{equation}
    \begin{equation}\label{eq:B}
         s_2 = B_{wm} = \frac{u_{\rm LES}}{u_{\tau_{wm}}} - \frac{1}{\kappa_{wm}} \ln \left( \frac{h_{wm} u_{\tau_{wm}}}{\nu} \right),
    \end{equation}
\end{subequations}
where $u_{\rm LES} = u(h_{wm})$ is the streamwise LES velocity taken at $h_{wm}$, and $u_{\tau_{wm}}=\sqrt{\tau_w/\rho}$ is the model-computed friction velocity. 
We place the matching location $h_{wm}$ at $dz$, i.e., between the first and the second off-wall grid points, for ease of computing the velocity derivative.
It may be worth noting that the implementation of RLWM is not straightforward. 
The present model requires the coupling of an RL library, here the Smarties library \cite{novati2019a} with the LES solver.

\begin{figure}
\centering
\begin{tabular}{cc}
\includegraphics[width=0.4\textwidth]{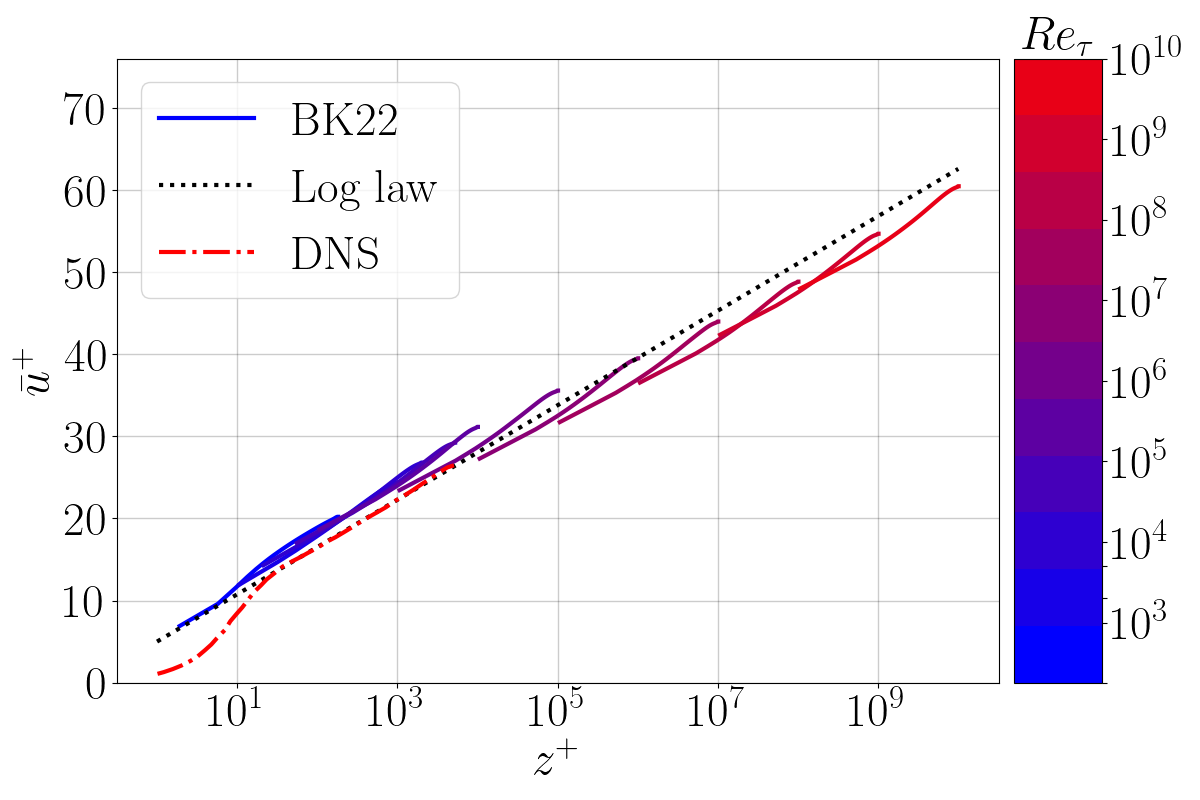}     & \includegraphics[width=0.4\textwidth]{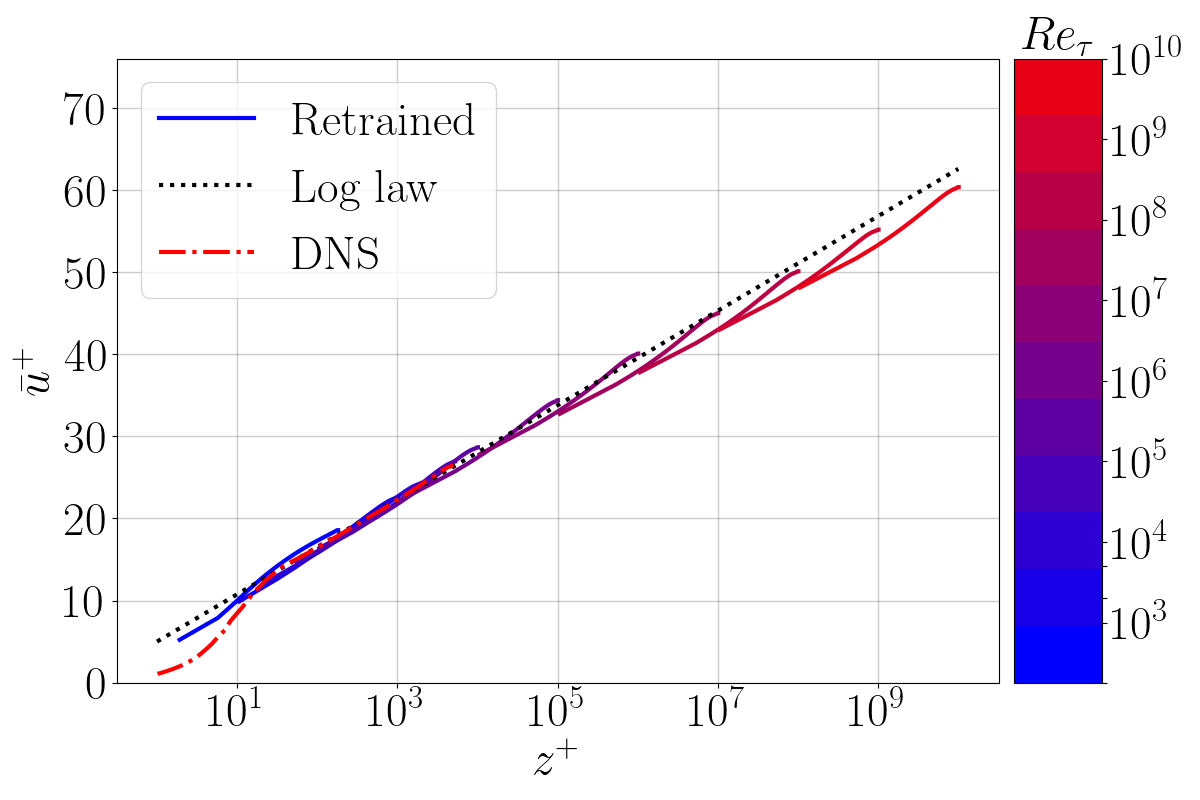} \\
   (a) BK22 tested in LESGO.  & (b) Retrained model in LESGO.
\end{tabular}
\caption{\label{fig:res1}Mean streamwise velocity $\bar{u}^+$ as a function of the wall-normal direction $z^+$ at 11 Reynolds numbers between $Re_\tau=180$ and $10^{10}$.
(a) BK22 tested in LESGO \cite{bae2022scientific}, (b) Retrained RLWM in LESGO. 
DNS result at $Re_\tau=5200$ is included for comparison purposes \cite{lee2015direct}.
The log law corresponds to $\kappa=0.4$ and $B=5$.}
\end{figure}

At first, BK22's results are recovered from this newly coupling between LESGO and the Smarties libraries. 
Training is performed on $Re_\tau=10^4$. 
Results are displayed in Figure \ref{fig:res1}. 
The same log-layer mismatch at large Reynolds numbers is observed for both models. 
At low Reynolds numbers, the log-layer mismatch is suppressed for the newly trained model. 
The models under comparison have identical states, rewards, and actions. 
The differences arise from the Reynolds number used in the training, which is slightly higher in our study, and from the specific RL settings. 
It is worth noting that the choice of solver may also contribute to differences since the BK22 model was initially trained using a solver with a second-order finite difference spatial scheme, whereas it is now tested in a pseudo-spectral code.
A comparison with the original paper \cite{bae2022scientific} can be found in Ref. \citenum{vadrot2022survey}.

\subsubsection{Reinforcement learning wall model, VYBA23}

The newly developed RLWM is based on the exact same RL algorithm as in Ref. \citenum{bae2022scientific}.
It is designed to address the limitations of the BK22 WM at high Reynolds numbers. 
The states in Equations \ref{eq:kappa} and \ref{eq:B} need to be corrected to achieve this. 
The limitations of the BK22 WM have been discussed in Ref. \citenum{vadrot2022survey}.
This issue arises from the rotation of the neutral line in the states-action map. 
The neutral line is defined as:
\begin{equation}\label{eq:neutral_line}
u^+_{\rm LES}-u_{\rm LL}^+ = 0.
\end{equation}
Here, $u_{LL}^+$ is the velocity obtained from the log law with $\kappa=0.4$ and $B=5$.
Both velocities are evaluated at the matching location $h_{wm}$. 
Equation \ref{eq:neutral_line} can be explicitly expressed from explicit quantities $\kappa$, $\kappa_{wm}$, $B$, $B_{wm}$ and $h_{wm}$ as: 
\begin{equation}
    \frac{1}{\kappa_{wm}}\ln(h_{wm}^+)+B_{wm} - \left(\frac{1}{\kappa}\ln(h_{wm}^+)+B \right)=0.
\end{equation}
For states located above this neutral line, the velocity is larger than the log-law value. 
Given these states, the RLWM should ideally generate an action $a_n>1$. 
By doing so, the wall-shear stress would increase, which is anticipated to result in a drop in the local velocity, thereby bringing down the velocity to the log-law value. 
On the contrary, given states below the neutral line, the RLWM should ideally generate an action $a_n<1$ to bring up the velocity to the log-law value. 
The reader will notice that RL models are amenable to physical interpretation. 
Visualizing the states-action map plot can help address the "black-box" criticism frequently attributed to ML models, as it enables a better comprehension of the learned policy of agents and its physical interpretation.

\begin{figure}
\centering
\begin{tabular}{cc}
 \includegraphics[width=0.435\textwidth]{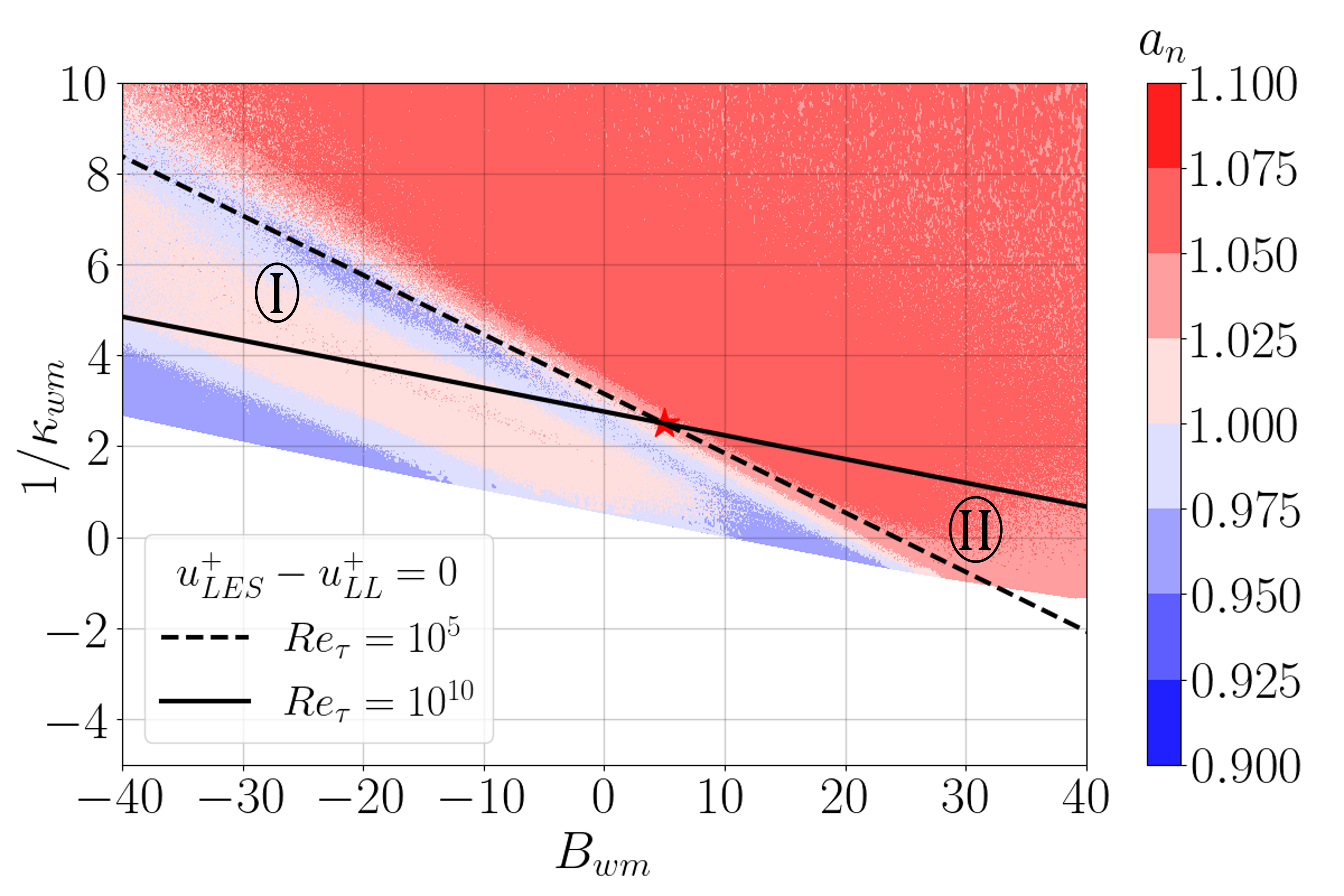} & \includegraphics[width=0.4\textwidth]{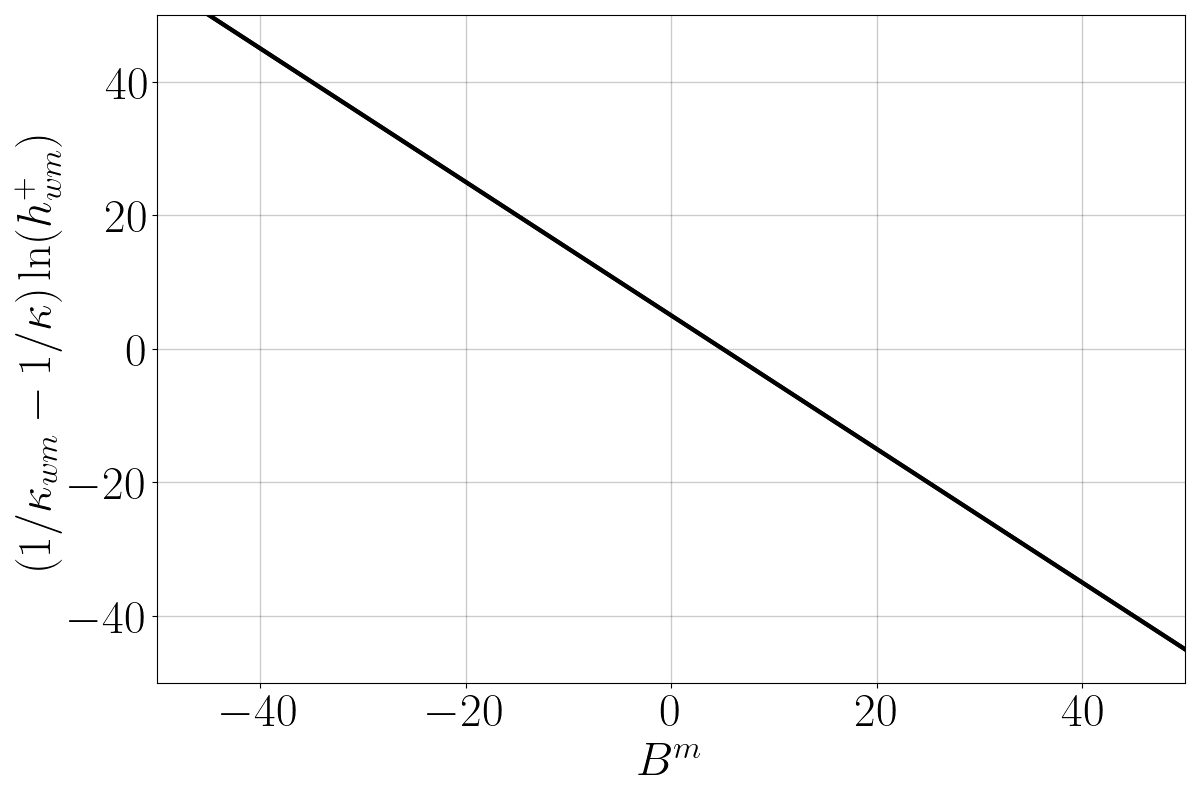} \\
 (a) States-action map for BK22 WM. & (b) States-action map with the new state $s_1$.
\end{tabular}
\caption{ \label{fig:new_norm} (a) States-action map at $Re_\tau=10^{10}$ for BK22 WM. The action contours show in red the increasing actions and in blue decreasing actions. The neutral lines are respectively plotted at $Re_\tau = 10^5$ and $10^{10}$. (b) States-action map with the new state $s_1 = \left( 1/{\kappa_{wm}}-1/{\kappa} \right) \ln(h_{wm}^+)$. The neutral line is identical for all $Re_\tau$.  } 
\end{figure}

Figure \ref{fig:new_norm} (a) displays the states-action map at $Re_\tau=10^{10}$ for BK22 WM. 
When increasing the Reynolds number, the neutral line rotates in counterclockwise and rotation saturates at large Reynolds numbers, explaining the saturating log-layer mismatch. 
This rotation is caused by the decrease of the slope of the neutral line $-1/(\ln(dz/\delta)+\ln(Re_\tau))$ for increasing $Re_\tau$ (see Ref. \citenum{vadrot2022survey} for further details).

To solve this issue, we propose to normalize the first state to remove the dependency on $h_{wm}^+$. 
If so, an agent trained for an arbitrary Reynolds number could be able to extrapolate to any unseen Reynolds number.
We propose the following expression for the first state $s_1$: 

\begin{equation}\label{eq:new_state}
    s_1 = \left( \frac{1}{\kappa_{wm}}-\frac{1}{\kappa} \right) \ln(h_{wm}^+).
\end{equation}
Now, the first state depends on the value of $\kappa$, but the neutral lines of the states-action map collapse into a single line whatever the Reynolds number is (see Figure \ref{fig:new_norm} (b)).  
The value of $\kappa$ in the log law is not universal and can depend on various factors such as the Reynolds number, surface roughness, and boundary layer type. 
Empirical studies have suggested values of $\kappa$ ranging from $0.36$ to $0.44$, with values around $0.4$ being commonly used as a default value \cite{nagib2008variations}.

\begin{table*}
\begin{ruledtabular}
\caption{\label{tab:detail} Details of VYBA23.}
\centering
\begin{tabular}{ccccccc}
WM & $\Delta a$ & $\Delta t$ &  $\Delta x$ & $\Delta y$& $N_{\rm agents}$ & Training $Re_\tau$  \\
\hline
{VYBA23} & $0.10$ & $10 dt$  & $3 dx$ & $3dy$ &$256$ &$10^4$\\
\end{tabular}
\end{ruledtabular}
\end{table*}

The parameters that are defined for agents in the context of multi-agents RLWM include the action range ($\Delta a$), where $a_n$ falls within the range $[1-\Delta a; 1+ \Delta a]$, the time-step between actions ($\Delta t$), and the horizontal ($\Delta x$) and vertical ($\Delta y$) distances between agents.
During the training of the VYBA23 WM, the parameters specified in Table \ref{tab:detail} were used.
The impact of these parameters on the flow will be examined in Section \ref{sec:result}.
The VYBA23 WM was trained from an LES at $Re_\tau=10^4$ using EWM, but this process was challenging and time-consuming. 
The training involved coupling the RL library, Smarties, which is coded in C++, with the Fortran-based CFD solver LESGO. 
Achieving proper convergence required a significant amount of trial and error with hundreds of simulations and millions of policy gradient updates. 
It is worth noting that some models may not converge, and several models with different reward functions were trained before settling on the VYBA23 model.

\section{\label{sec:result} Results}

We trained a new RLWM, namely VYBA23, using the new expression for the first state (Equation \ref{eq:new_state}) and based on parameters from Table \ref{tab:detail}. 
In this section, we evaluate this model. 
The BK22 WM \cite{bae2022scientific} and the EWM \cite{kawai2012wall} are sometimes show for comparison.  

\subsection{The effect of reinforcement learning wall models}

\begin{figure}
\centering
\begin{tabular}{ccc}
 \includegraphics[width=0.3\textwidth]{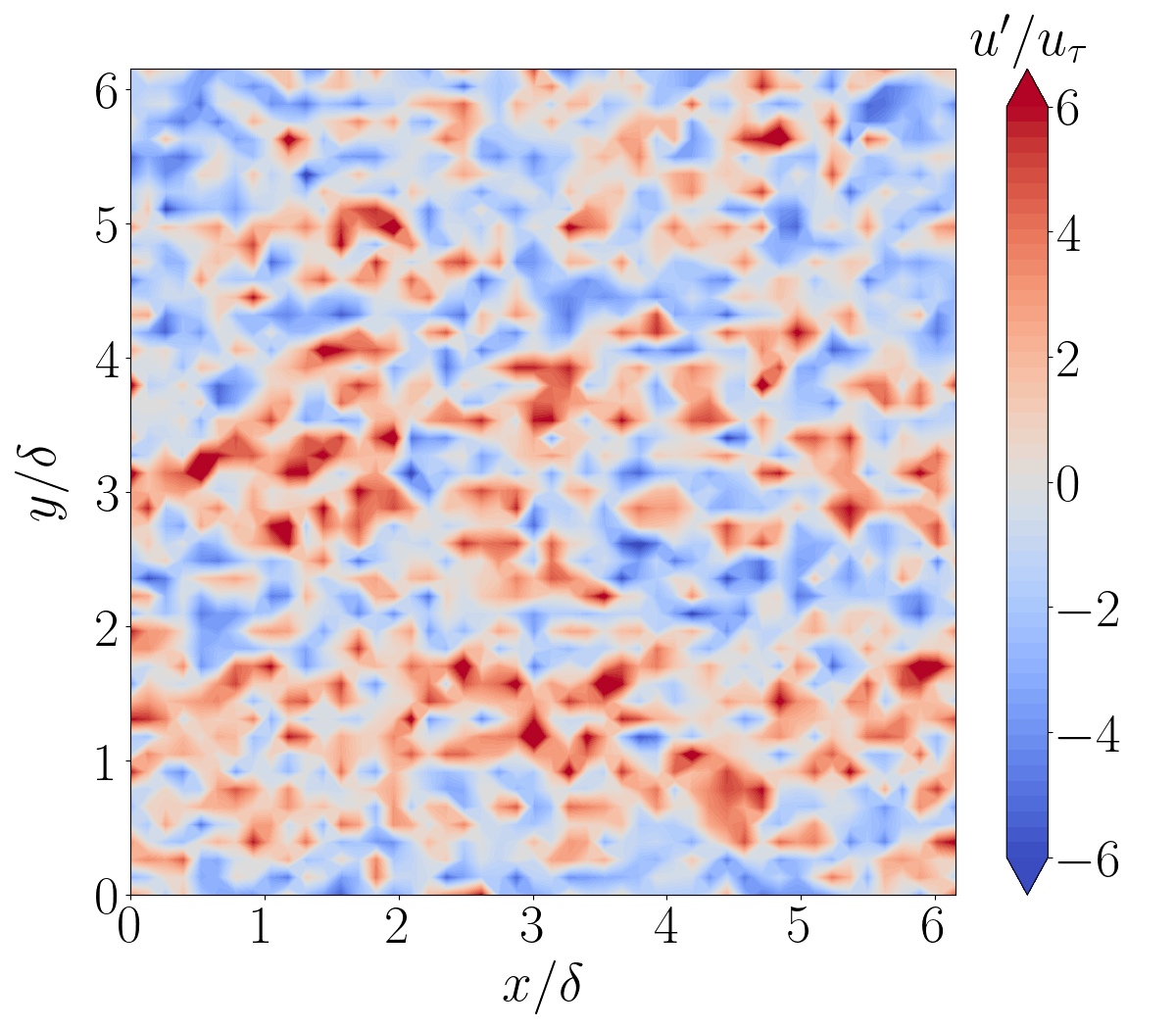}    &   \includegraphics[width=0.3\textwidth]{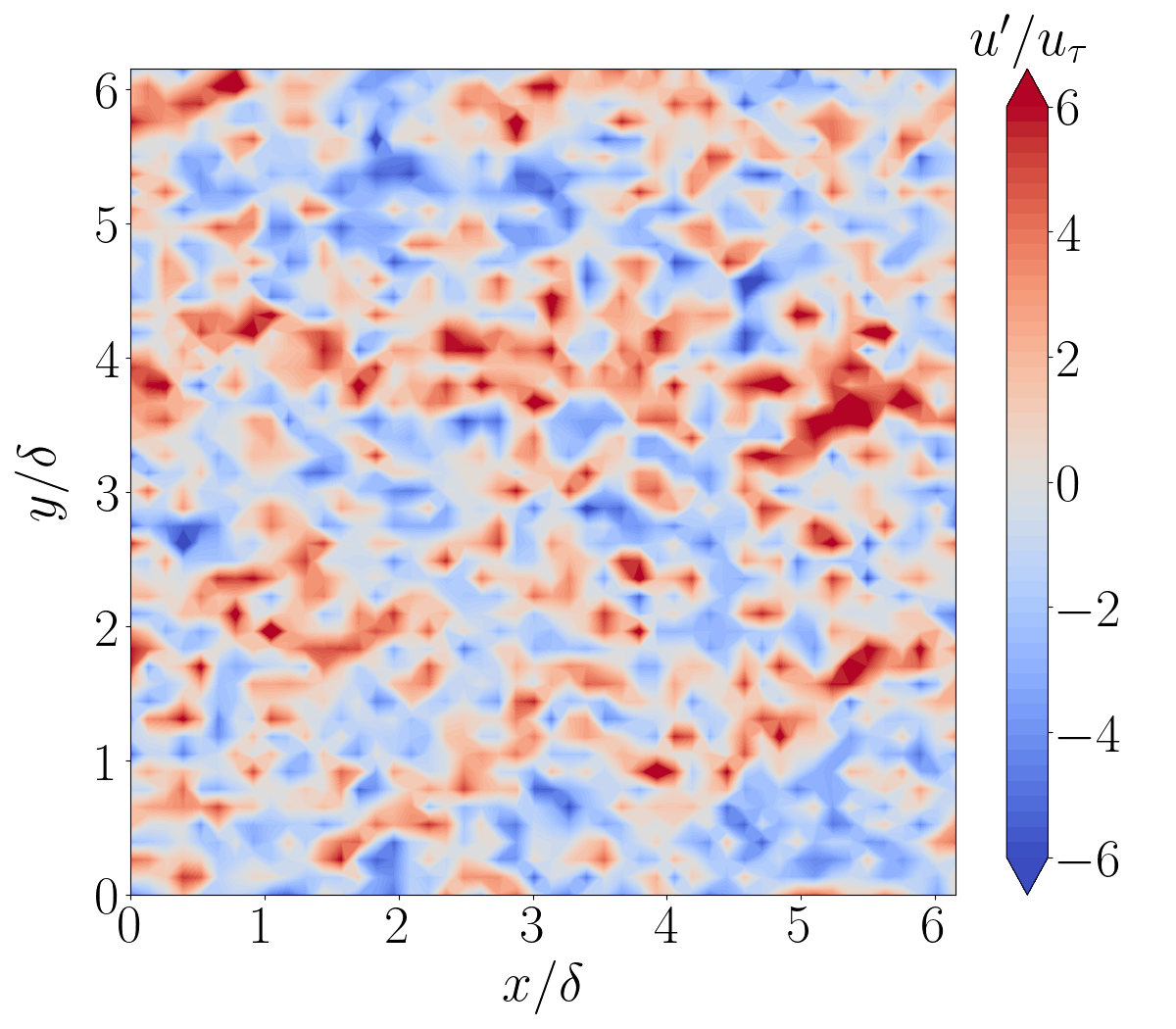} &\includegraphics[width=0.3\textwidth]{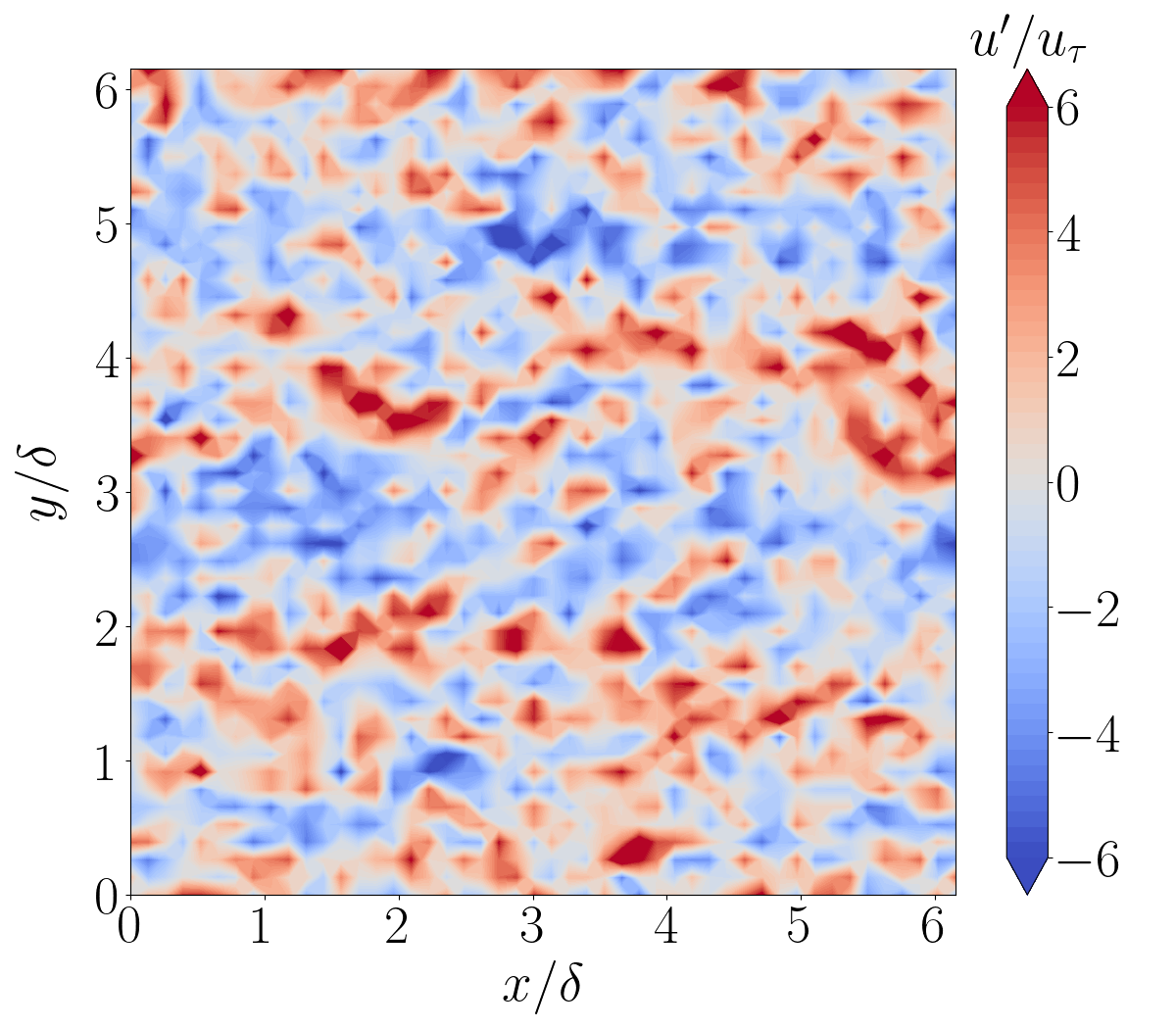} \\
 (a) EWM. & (b) BK22. & (c) VYBA23.\\
\end{tabular}
\caption{\label{fig:planxy1} Contours of the fluctuating streamwise velocity in the $x-y$ plane at the first off-wall grid point $z=dz/2$.
The flow is at $Re=10^5$.
(a) EWM \cite{kawai2012wall}, (b) BK22 \cite{bae2022scientific}, (c) VYBA23. 
The grid resolution is $N_x \times N_y \times N_z = 48^3$.} 
\end{figure}

\begin{figure}
\centering
\begin{tabular}{ccc}
\includegraphics[width=0.3\textwidth]{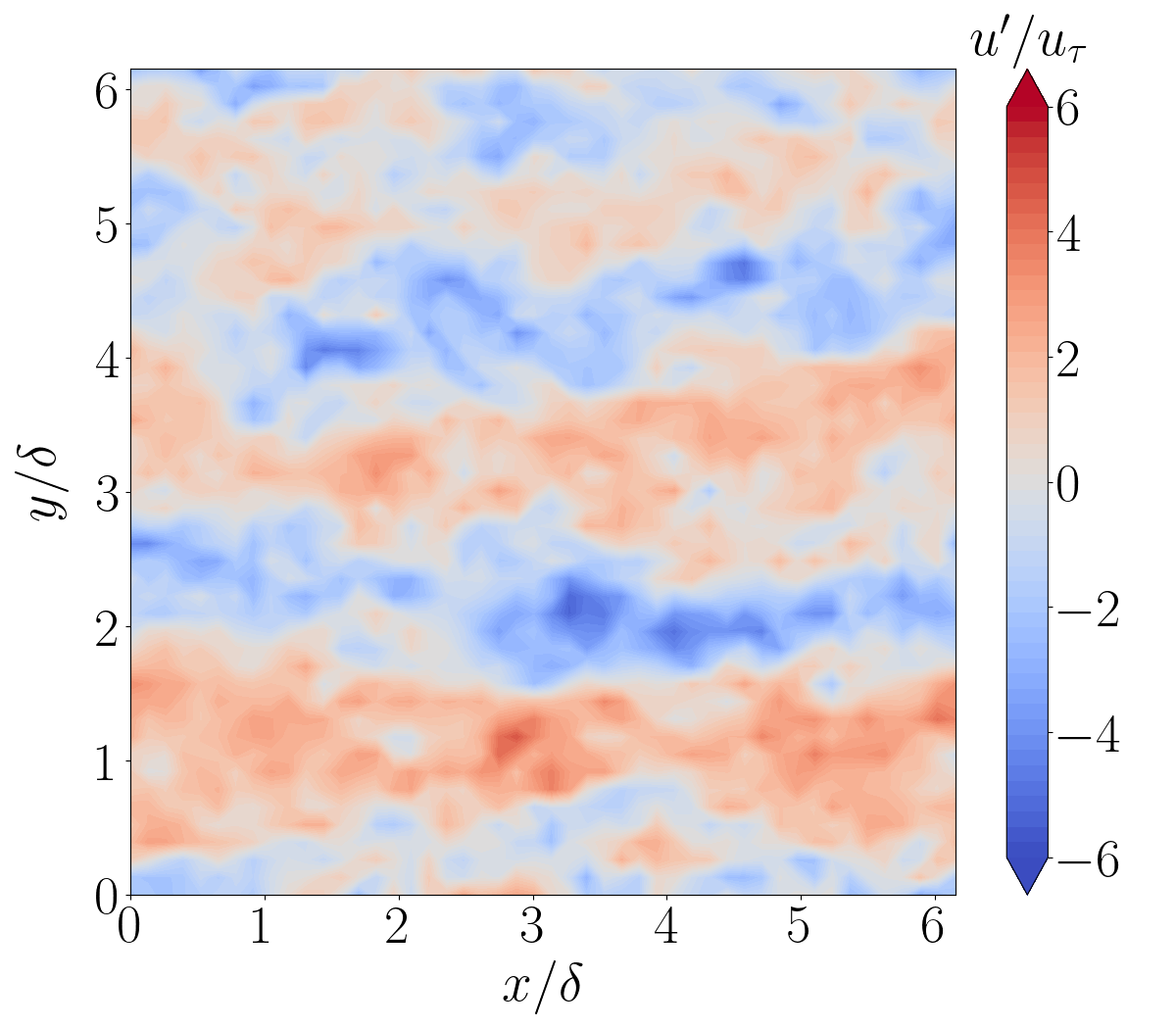}    &   \includegraphics[width=0.3\textwidth]{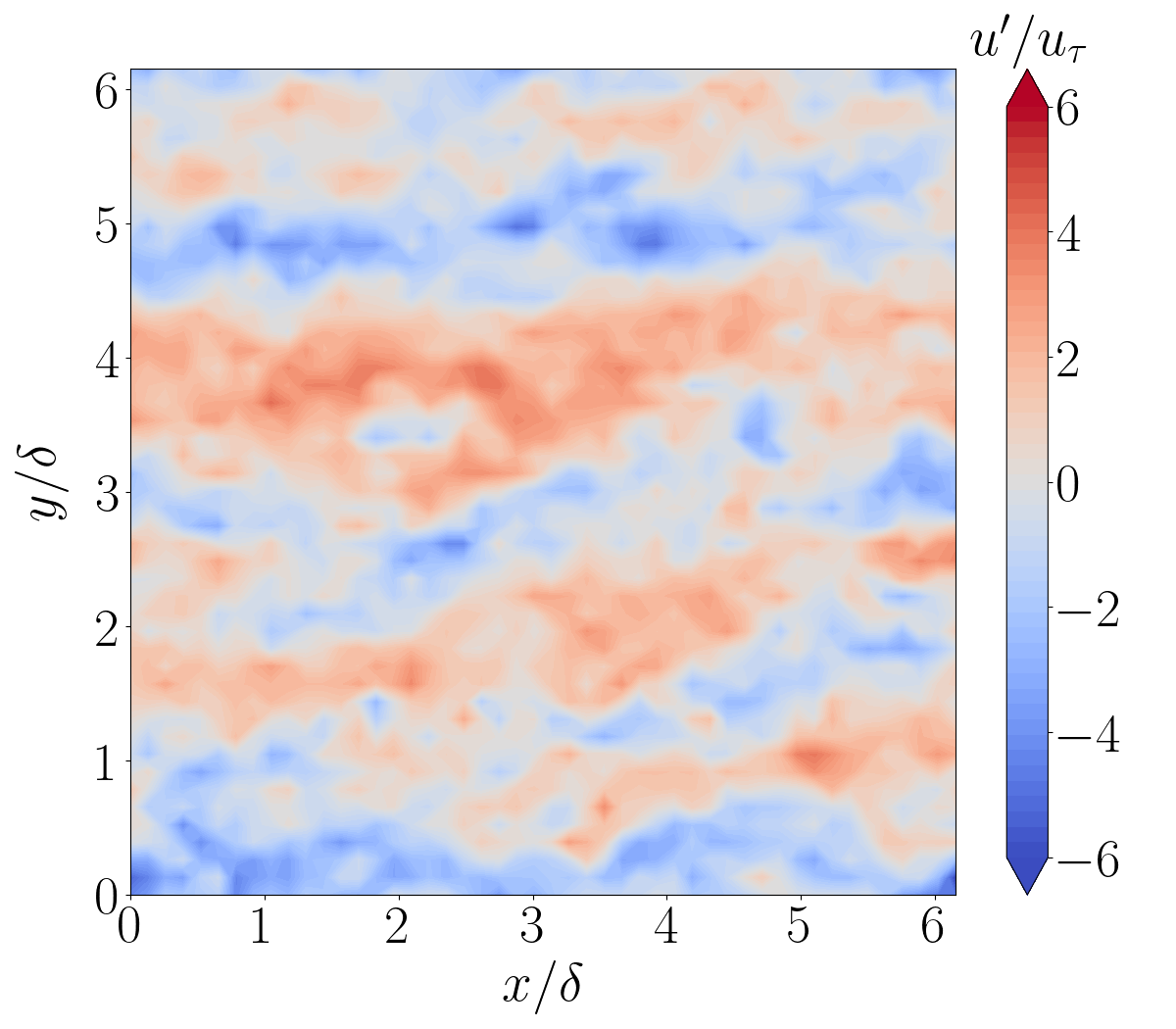} &\includegraphics[width=0.3\textwidth]{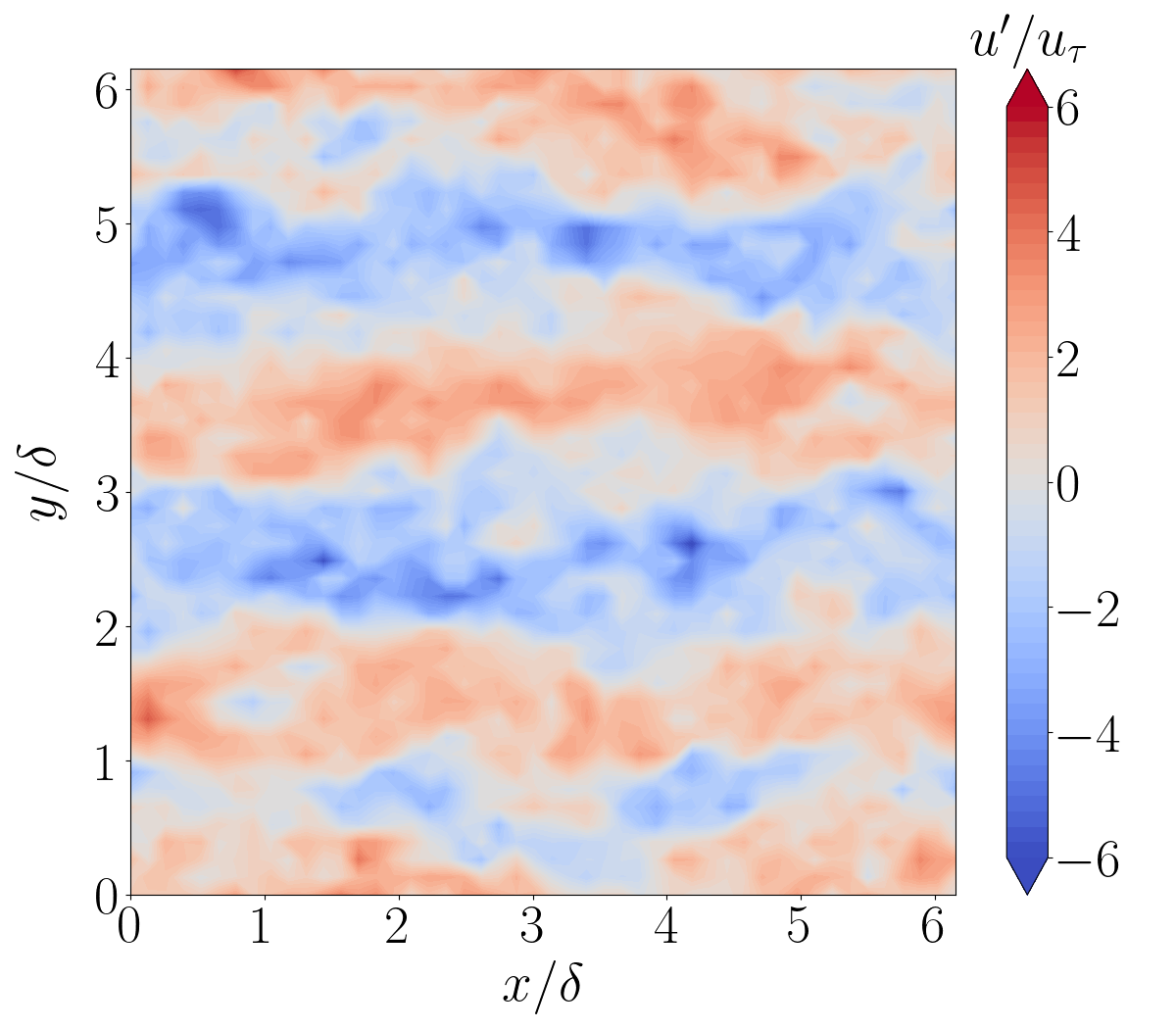} \\
 (a) EWM. & (b) BK22. & (c) VYBA23.\\
\end{tabular}
\caption{\label{fig:planxy2}Same as Figure \ref{fig:planxy1} but at $z/\delta=0.5$.} 
\end{figure}

 Figures \ref{fig:planxy1} and \ref{fig:planxy2} show instantaneous snapshots of the contours of streamwise velocity fluctuations at respectively the first off-grid point wall and at $z/\delta=0.5$. 
 The friction Reynolds number is equal to $10^5$. The flow fields are alike between EWM \cite{kawai2012wall}, BK22 \cite{bae2022scientific} and VYBA23. 
 The flow is composed of high-intense small-scale turbulent structures in the near-wall region and of large-scale streaks at $z/\delta=0.5$. 
 For all the three WMs, both intensity and scale-sizes are similar.

\begin{figure}
\centering
{\includegraphics[width=0.4\textwidth]{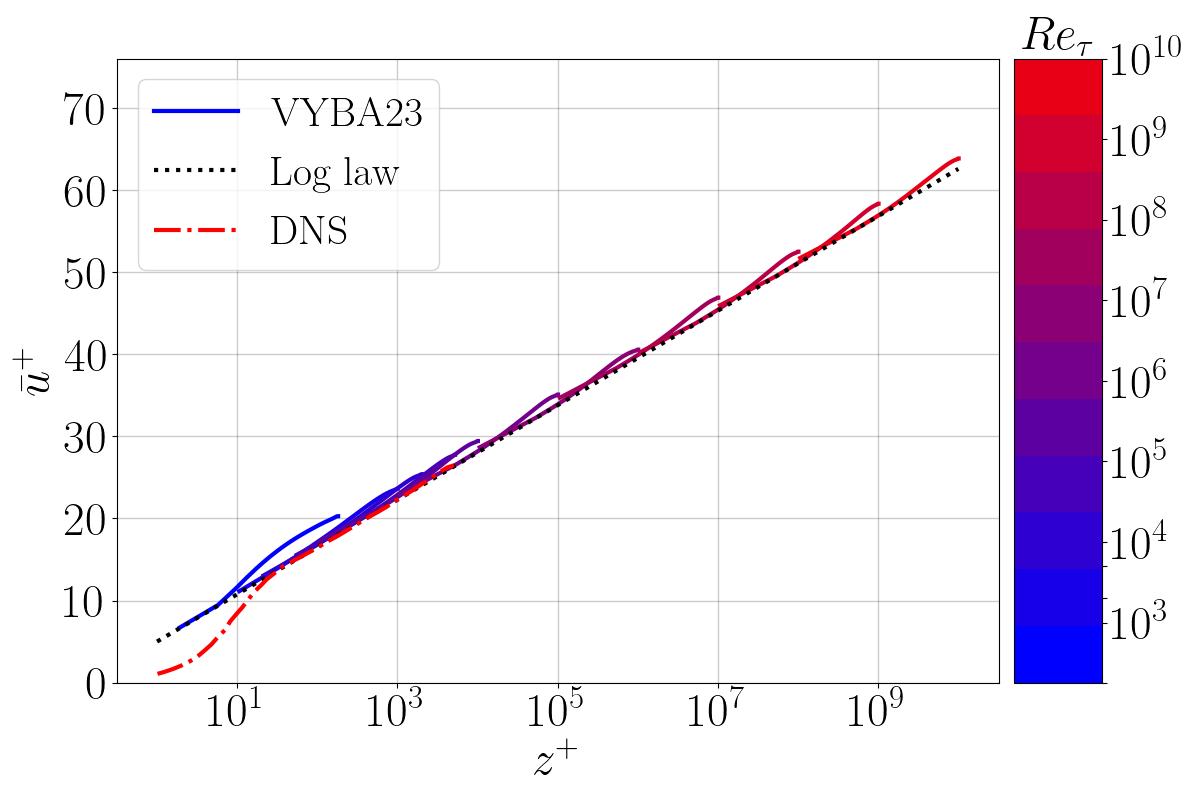}}
\caption{\label{fig:res} Mean streamwise velocity $\bar{u}^+$ as a function of the wall-normal direction $z^+$ at 11 Reynolds numbers between $Re_\tau=180$ and $10^{10}$ for VYBA23 WM. 
DNS result at $Re_\tau=5200$ is included for comparison purposes \cite{lee2015direct}.
The log law corresponds to $\kappa=0.4$ and $B=5$. }
\end{figure}

 Figure \ref{fig:res} displays the mean velocity profiles. 
 The VYBA23 WM captures properly the log law when the matching location $h_{wm}$ is in the log layer (i.e. at $Re_\tau = 180$, the prediction is above the DNS profile). 
 The VYBA23 WM outperforms the BK22 WM \cite{bae2022scientific} (Figure \ref{fig:res1}), providing an accurate prediction of the log law outside of its training range ($Re_\tau=10^4$), in high Reynolds numbers channel flows (up to $Re_\tau= 10^{10}$).          

 \begin{figure}
\centering
\begin{tabular}{ccc}
\includegraphics[width=0.32\textwidth]{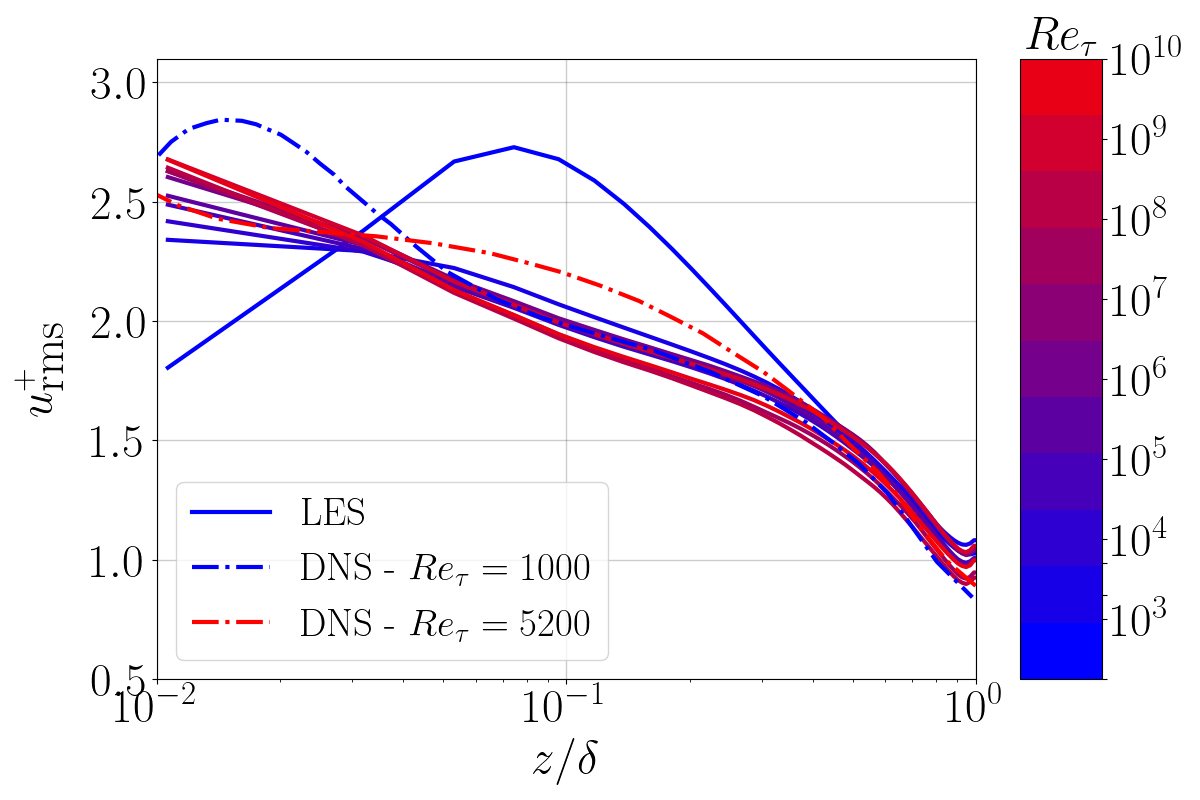} & \includegraphics[width=0.32\textwidth]{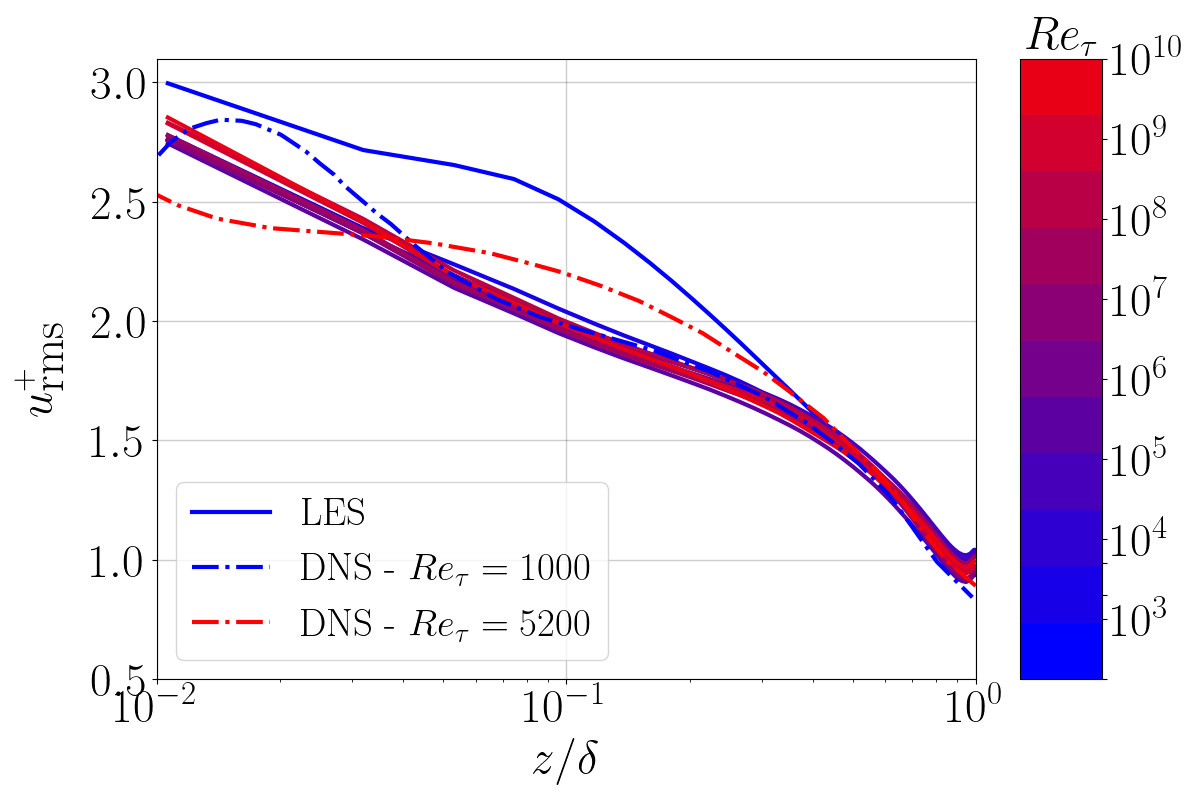} & \includegraphics[width=0.32\textwidth]{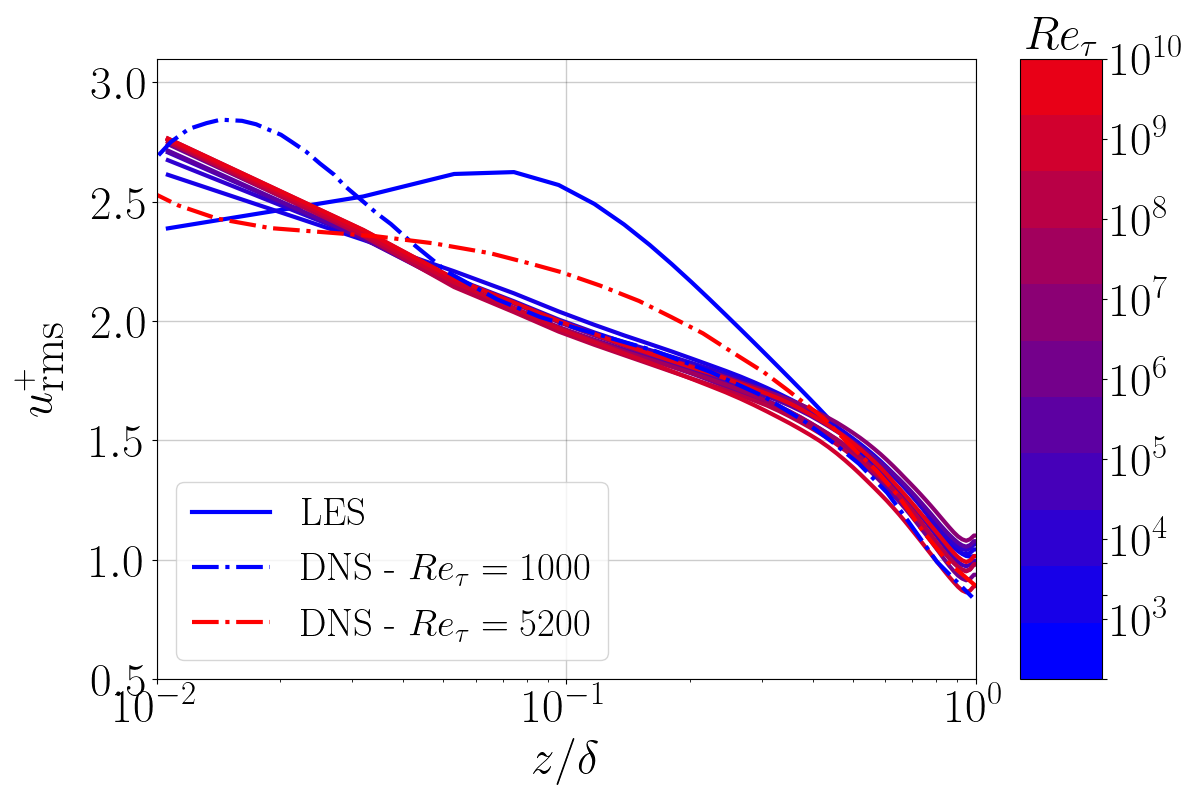} \\
 (a) EWM. & (b) BK22. & (c) VYBA23. \\ 
 \end{tabular}
\caption{\label{fig:rms} Root mean square of the streamwise velocity fluctuation $u_\textrm{rms}^+$ as a function of the wall-normal coordinate $z/\delta$ plotted for between $Re_\tau=180$ and $10^{10}$. (a) EWM \cite{kawai2012wall}, (b)  BK22 \cite{bae2022scientific}, (c) VYBA23. DNS data at $Re_\tau=1000$ and $Re_\tau=5200$ are included for comparison\cite{graham2016web,lee2015direct}.} 
\end{figure}

Figure \ref{fig:rms} displays the root mean square (rms) value of the velocity fluctuations $u^+_{\rm rms}$ for EWM, BK22 and VYBA23 WMs. 
\blue
The recent DNS and experimental studies of boundary-layer flows at high Reynolds numbers have revealed an outer peak in $u_{\rm rms}^{+^2}$ in addition to the inner peak \cite{marusic2013logarithmic,hultmark2012turbulent}.
%Here, neither the inner nor outer peaks can be observed, and the reasons for this are as follows. 
Here, the inner peak cannot be observed, and the outer peak is barely noticeable.
The explanations for this are provided below.
The bump caused by the outer layer peak is visible at $z/\delta\approx 0.1$. 
However, none of the WMs capture this bump, likely due to insufficient resolution to accurately represent the outer peak.
The inner peak is located at around $z^+\approx 15$ within the buffer layer, while the outer peak emerges at approximately $z^+\approx 3\sqrt{Re_\tau}$ for high Reynolds numbers\cite{marusic2013logarithmic}, corresponding to $z/\delta=3/\sqrt{Re_\tau}$. 
For $Re_\tau= 10^4$, the outer peak is at $z/\delta\approx 0.03$. 
As the Reynolds number increases, the outer peak moves closer to the wall in terms of $z/\delta$. 
Since the grid resolution in WMLES is coarse, and a typical WMLES does not resolve either the inner or outer peaks. 
From the outer peak towards the channel center, the streamwise velocity rms scales as:

\begin{equation}
u^{+^2}_{\rm rms}\sim \ln(\delta/z).
\end{equation}
This scaling is independent of the Reynolds number, which is why the WMLES-predicted $u_{\rm rms}^{+}$ collapses at sufficiently high Reynolds numbers. 
The curve at $Re_\tau=180$ differs from the others because the Reynolds number is too low. 
%Apart from that, all curves converge into a single distribution, in line with the expectations of Townsend's attached eddy model \cite{townsend1980structure}.
\black

\begin{figure}
\centering
{\includegraphics[width=0.4\textwidth]{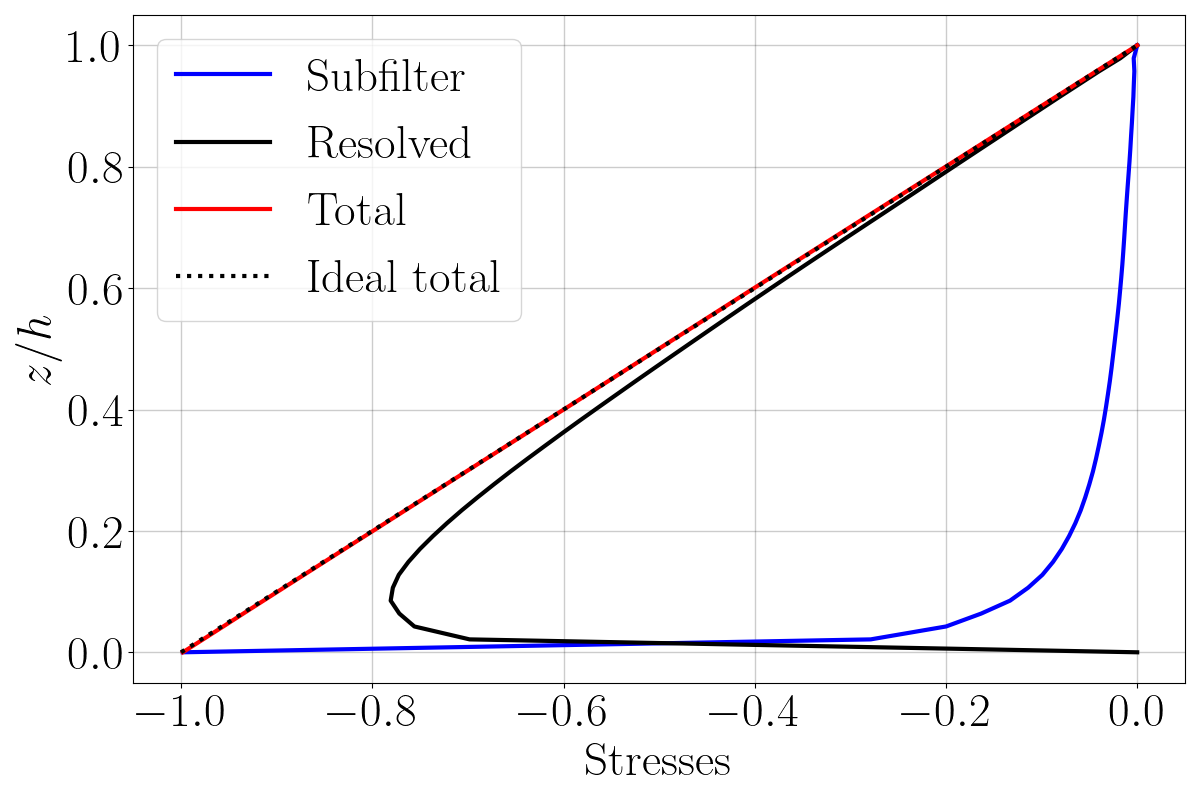}}
\caption{\label{fig:stress} Vertical profiles of the normalized total and partial shear stress for $Re_\tau=10^8$ for VYBA23 WM. }
\end{figure}

It is important to note that in this study, each LES was carefully checked for statistical convergence by examining the convergence of kinetic energy and the stress balance. 
It is well-established that once equilibrium is reached, the total stress, which is the sum of the filter (modeled) and resolved stresses, becomes a linear function of the wall-normal distance. 
Figure \ref{fig:stress} demonstrates the successful convergence of the total stress for $Re_\tau=10^8$. 
The figure shows that the total stress obtained almost perfectly overlaps with the ideal total stress.

\subsection{Further results}
\label{sec:further}

\begin{table*}
\begin{ruledtabular}
\caption{\label{tab:overview} Test of the VYBA23 WM with different running settings. }
\centering
\begin{tabular}{ccccc}

 $\Delta a$ & $\Delta t$ &  $\Delta x$ & $\Delta y$& $N_{\rm agents}$ \\
\hline
\\
\hline
 $0.10$ & $10 dt$  & $3 dx$ & $3dy$ &$256$\\
 \hline 
  $0.10$ & $10 dt$  & $3 dx$ & $6dy$ &$128$\\
   \hline
  $0.10$ & $10 dt$  & $6 dx$ & $6dy$ &$64$\\ 
  \hline
  \\
 \hline
  $0.05$ & $10 dt$  & $3 dx$ & $3dy$ &$256$\\
   \hline
  $0.025$ & $10 dt$  & $3 dx$ & $3dy$ &$256$\\ 
\hline
  \\
\hline
  $0.10$ & $dt$  & $3 dx$ & $3dy$ &$256$\\
   \hline
  $0.10$ & $5 dt$  & $3 dx$ & $3dy$ &$256$\\
   \hline
  $0.10$ & $20 dt$  & $3 dx$ & $3dy$ &$256$\\ 

%  & $0.01$ & $dt$  & $3 dx$ & $3dy$ &$256$\\
% & $0.05$ & $dt$  & $3 dx$ & $3dy$ &$256$\\
% & $0.20$ & $dt$  & $3 dx$ & $3dy$ &$256$\\

\end{tabular}
\end{ruledtabular}
\end{table*}

\begin{figure}
\centering
\begin{tabular}{cc}
\includegraphics[width=0.4\textwidth]{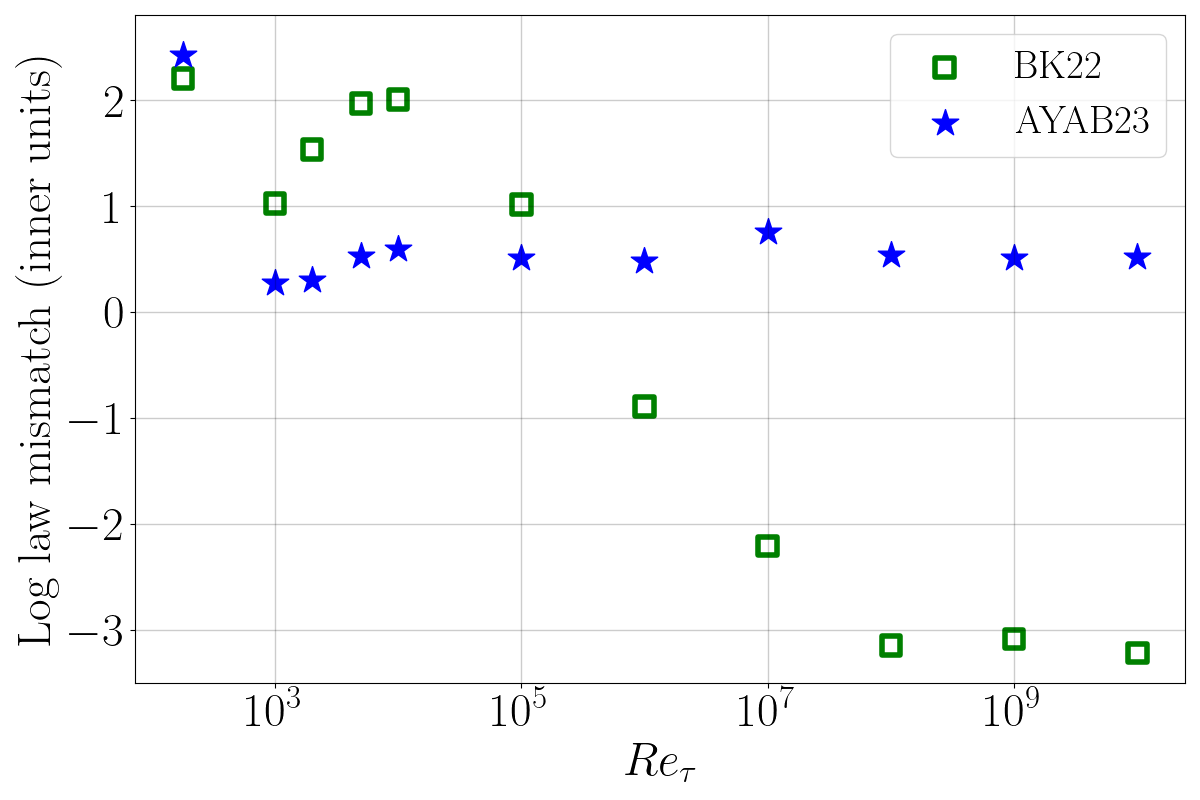}   &  \includegraphics[width=0.4\textwidth]{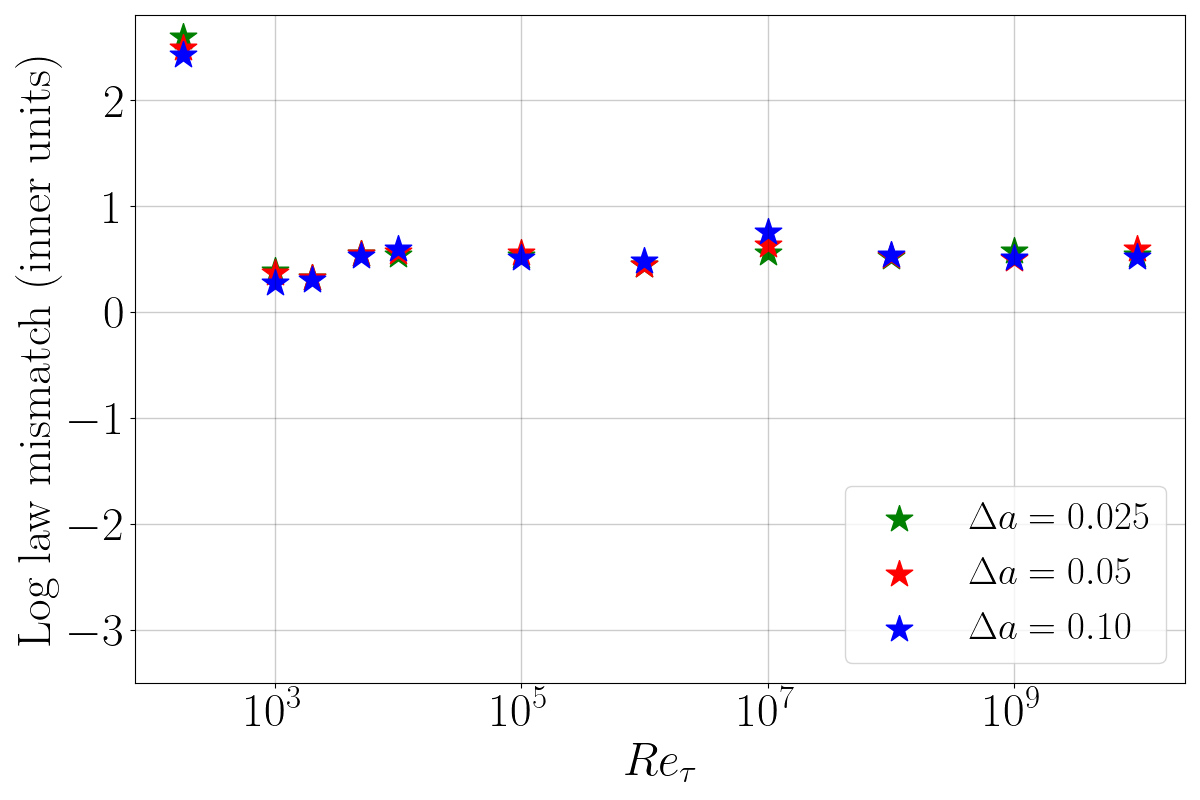}\\
 (a) BK22. & (b) Effect of $\Delta a$. \\ 
 \includegraphics[width=0.4\textwidth]{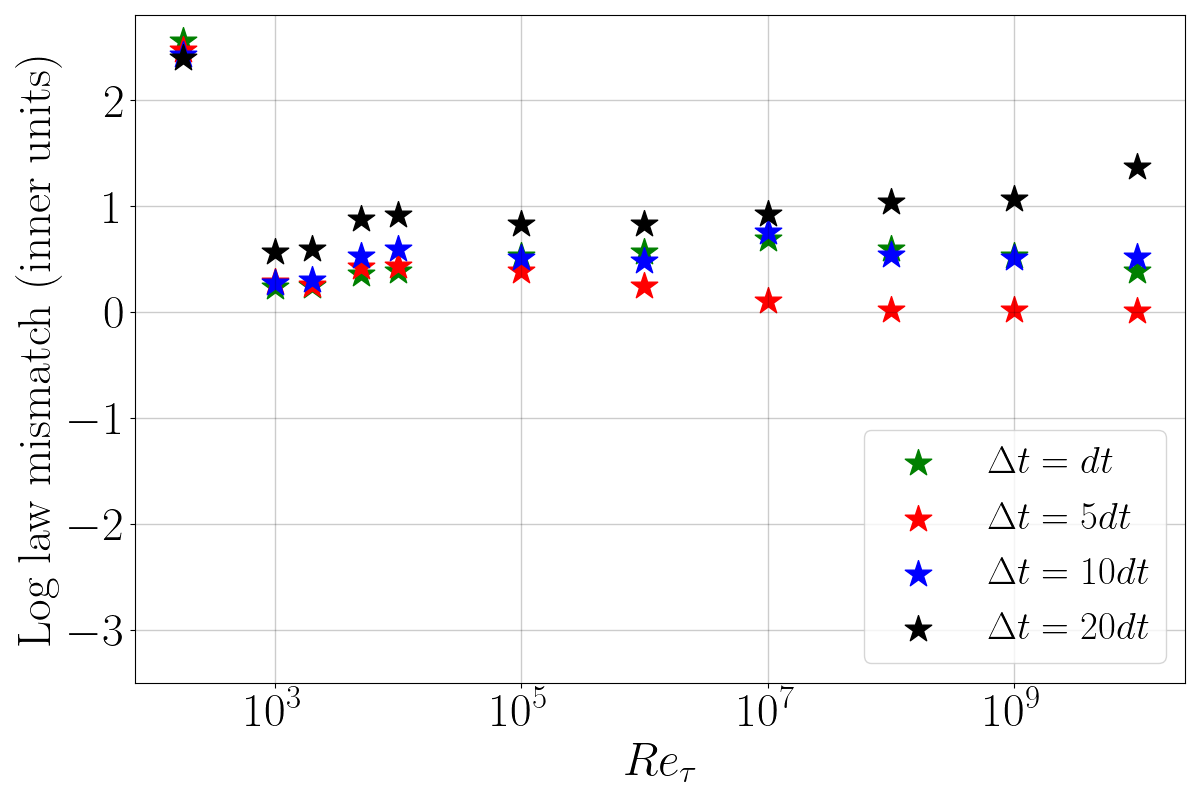}   &  \includegraphics[width=0.4\textwidth]{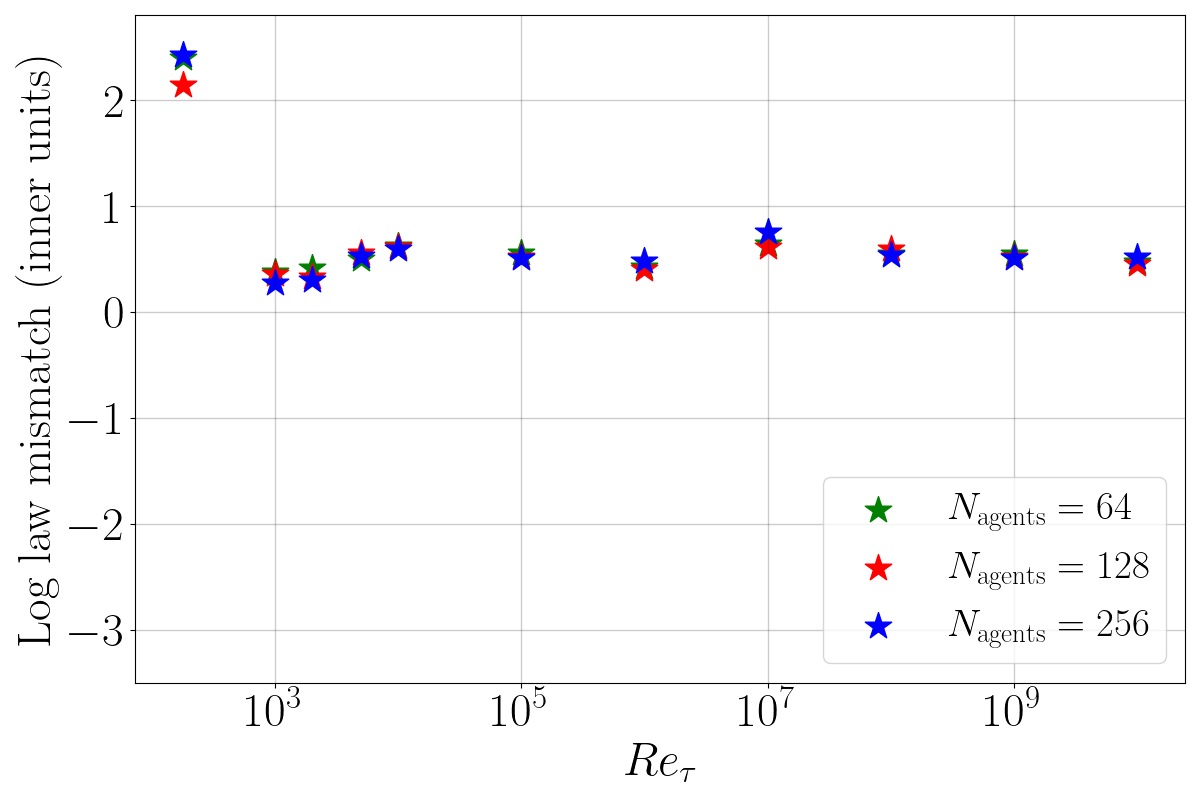}\\
 (c) Effect of $\Delta t$. & (d) Effect of $N_{\rm agents}$. \\ 
 \end{tabular}
\caption{\label{fig:mismatch_LL} Log-layer mismatch in inner units as a function of the friction Reynolds number $Re_\tau$ for BK22 \cite{bae2022scientific} and VYBA23. 
The agents' parameters are tested individually by changing one parameter at a time while keeping the others at their baseline configuration ($\Delta a = 0.10$, $\Delta t= 10 dt$ and $N_{\rm agents}=256$, see Table \ref{tab:detail}).
The baseline log law is $\ln(z^+)/\kappa+B$ with $\kappa=0.4$ and $B=5$.} 
\end{figure}

\begin{figure}
\centering
\begin{tabular}{cc}
\includegraphics[width=0.4\textwidth]{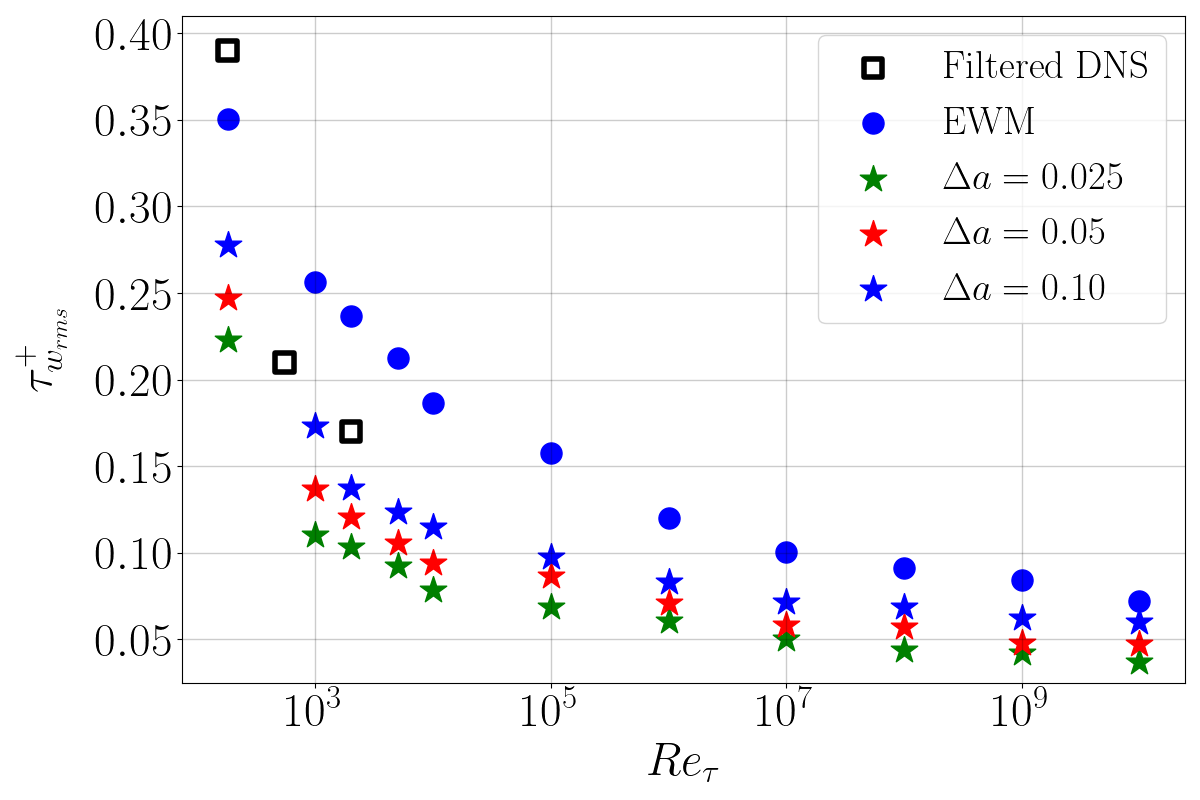}  &  \includegraphics[width=0.4\textwidth]{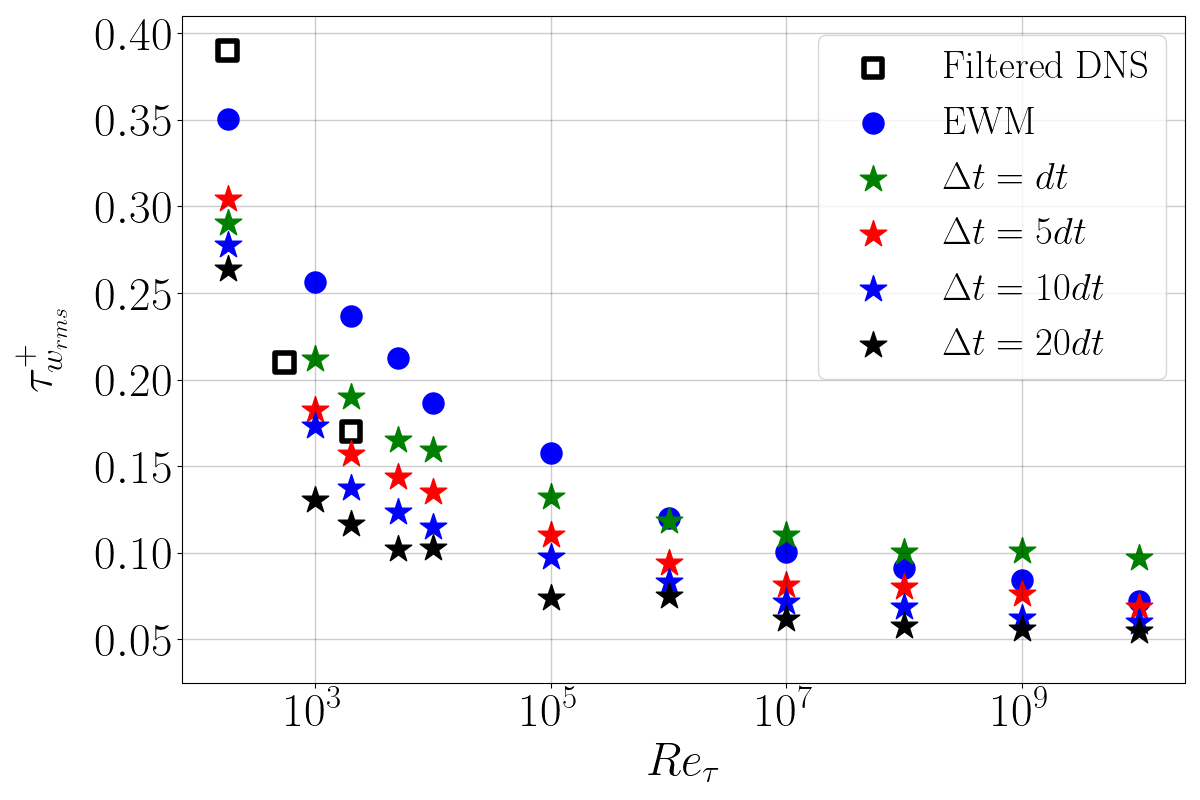}\\
 (a) Effect of $\Delta a$. & (b) Effect of $\Delta t$. \\ 
 \end{tabular}
\includegraphics[width=0.4\textwidth]{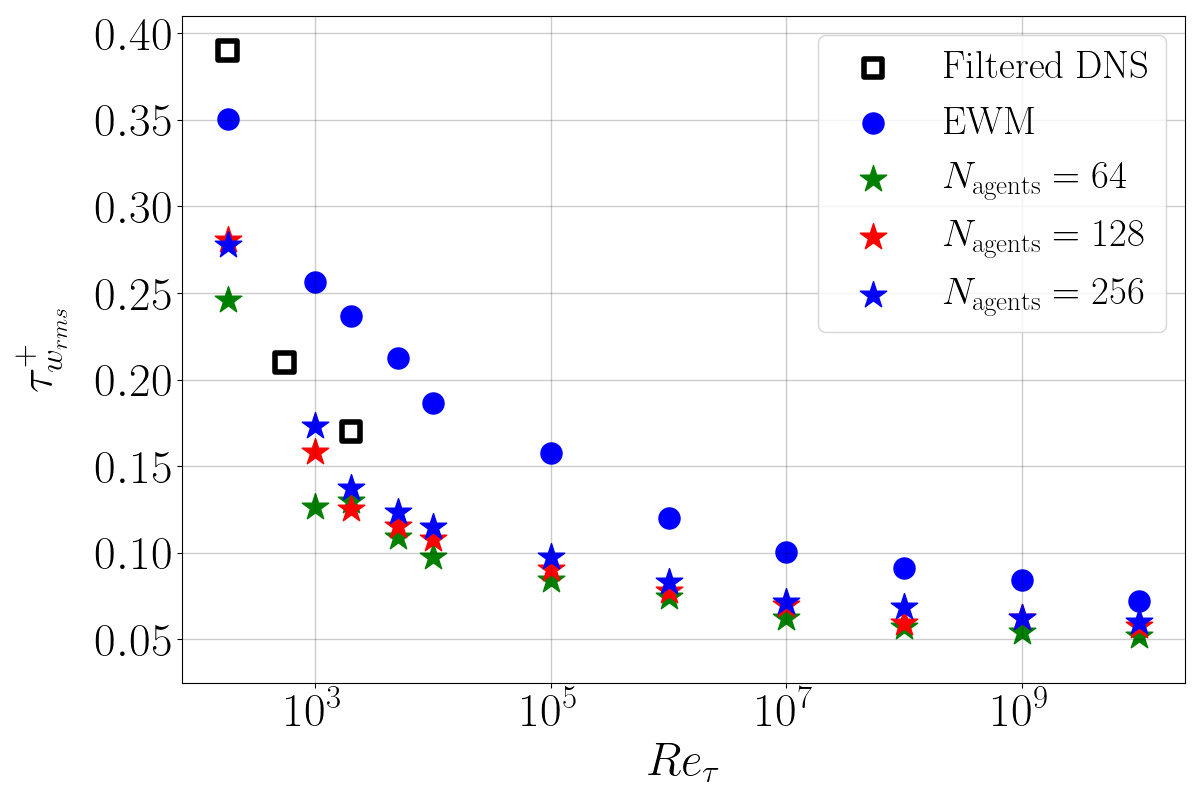}\\
(c) Effect of $N_{\rm agents}$.
\caption{\label{fig:taurms} Evolution of wall-shear stress fluctuations for VYBA23, EWM \cite{kawai2012wall} and filtered DNS \cite{yang2017log}. DNS are filtered with a top-hat filter. 
The agents' parameters are tested individually by changing one parameter at a time while keeping the others at their baseline configuration ($\Delta a = 0.10$, $\Delta t= 10 dt$ and $N_{\rm agents}=256$, see Table \ref{tab:detail}).} 
\end{figure}

In order to assess the impact of agents' parameters on the results, various combinations were tested by varying the agents' spacing ($\Delta x$ and $\Delta y$), the range of actions ($\Delta a$), and the time-step ($\Delta t$). 
The tested combinations are listed in Table \ref{tab:overview}.
Figure \ref{fig:mismatch_LL} illustrates the influence of $\Delta a$, $\Delta t$, and $N_{\rm agents}$ on the log-layer mismatch. 
A 5\% error in the predicted wall-shear stress is expected due to a $2.5\%$ uncertainty in the von Kármán constant $\kappa$ \cite{yang2019predictive}.
This translates to about 1-2 plus units in the context of log-layer mismatch.
The log-layer mismatch problem that was identified in the BK22 WM \cite{vadrot2022survey} has been addressed, with most of the parameters of the agents having limited impact on the mean velocity distribution. 
The only exception being when the time-step between actions becomes too large ($\Delta t > 10dt$). 
That strengthens the reliability of RLWM methods, where the mean field is largely unchanged by the agents' parameters. 
However, the parameters may still affect the higher-order statistics of the flow, particularly the fluctuations in the wall-shear stress. 
RLWM adjusts the wall-shear stress to match its true mean value (as shown in Equation \ref{eq:true_tauw}), leading to varying levels of fluctuations depending on the choice of agents' parameters.
It is acknowledged that fluctuations of the wall-shear stress can have a significant impact on the flow \cite{diaz2017wall}.
The selection of agents' parameters provides a great deal of flexibility (as seen in Figure \ref{fig:taurms}).

The fluctuations of the wall-shear stress decrease as the friction Reynolds number increases.
This can be explained by re-writing Equation \ref{eq:wall-shear} as:

\begin{equation}
    \tau_w = \rho u_\tau^2 = \rho \left[ \frac{\kappa \tilde{U}_{\rm LES}}{\ln(h_{wm}^+) + B}\right ].
\end{equation}
Wall-shear fluctuations depend on $1/(\ln(h_{wm}^+ +B)$ that causes the wall-shear fluctuations decrease for increasing Reynolds numbers.
The range and time-step of actions are the most influential parameters, with fluctuations increasing with range and decreasing with time-step. 
The impact of agents' spacing on the wall-shear stress fluctuations is minimal, as states are likely to be similar between two agents and adding an agent between them would not result in significant differences compared to the interpolation of the wall-shear stress.
The optimal setting to match filtered DNS based on wall-shear stress fluctuations results is $(\Delta a; \Delta t; \Delta x; \Delta y)= (0.10; 5dt;3dx;3dy)$. 
However, it is important to note that this choice may be influenced by the numerical schemes used.

\section{\label{sec:analysis} Analysis}

 \begin{figure}
\centering
\begin{tabular}{cc}
\includegraphics[width=0.4\textwidth]{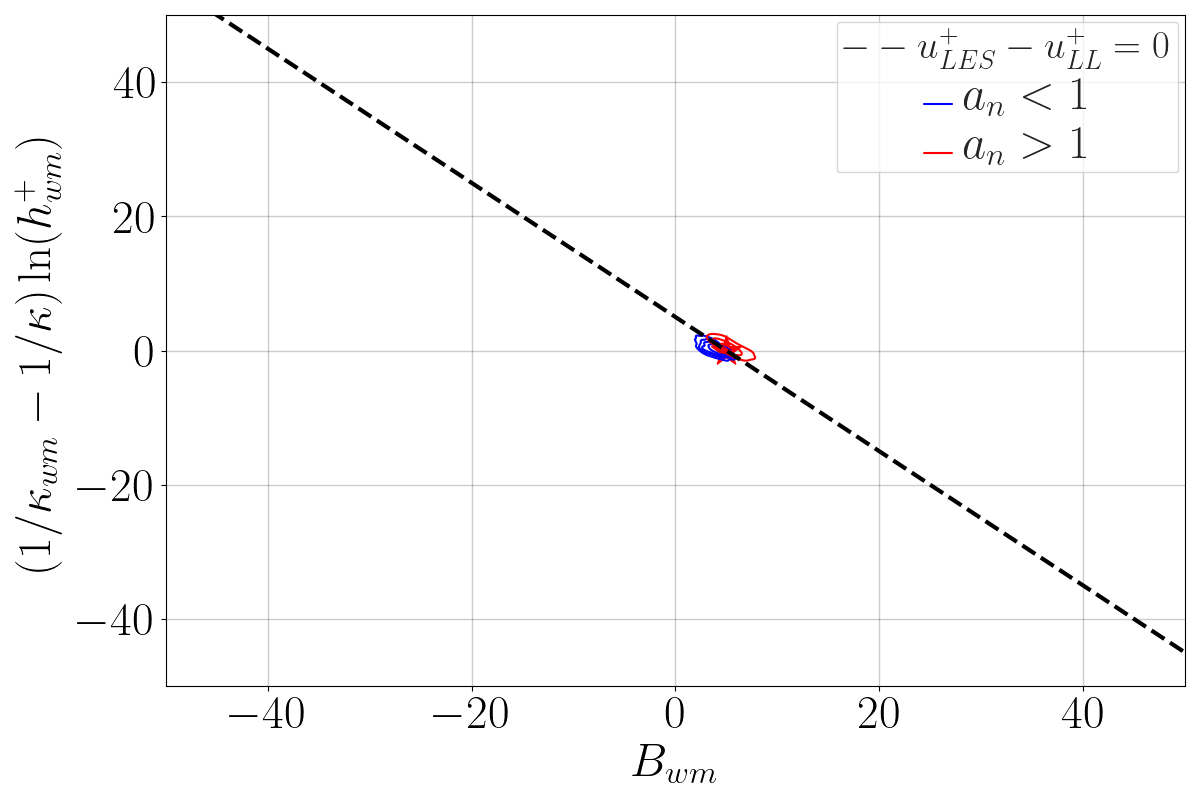}  &  \includegraphics[width=0.4\textwidth]{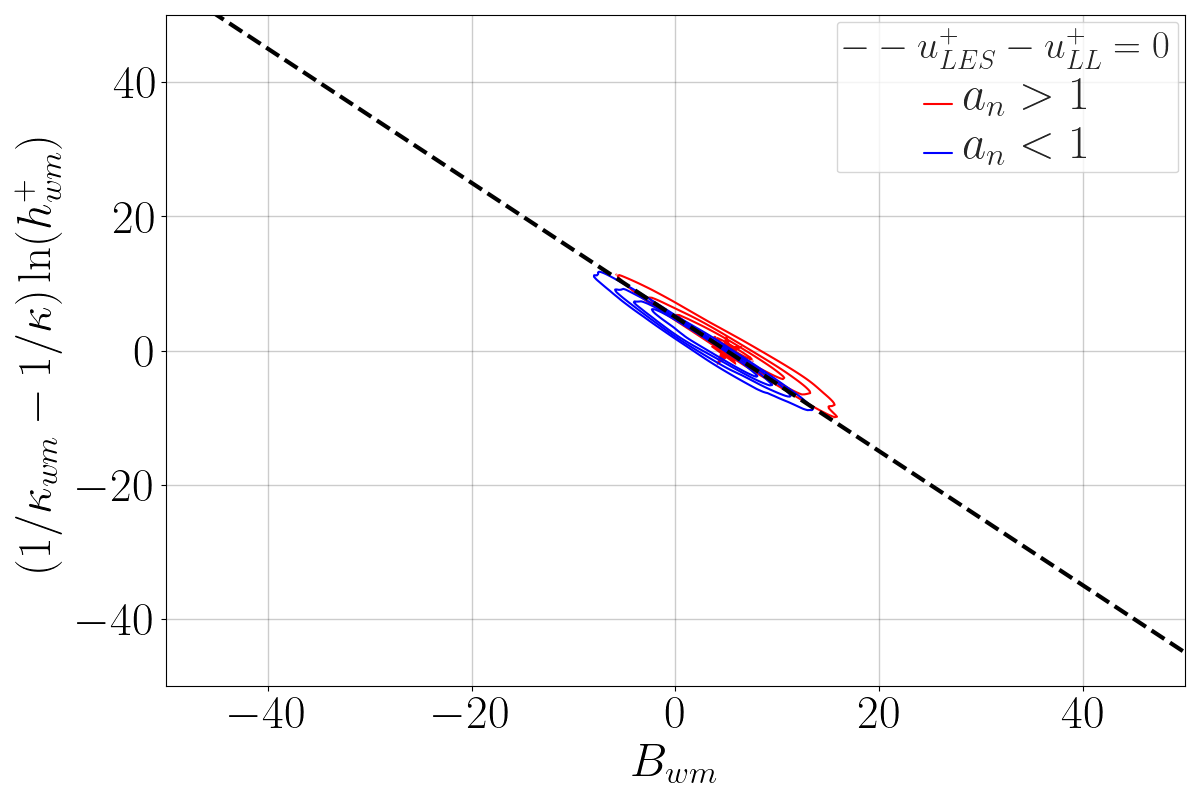}\\
 (a) $Re_\tau=180$. & (b) $Re_\tau=10^{3}$. \\ 
\includegraphics[width=0.4\textwidth]{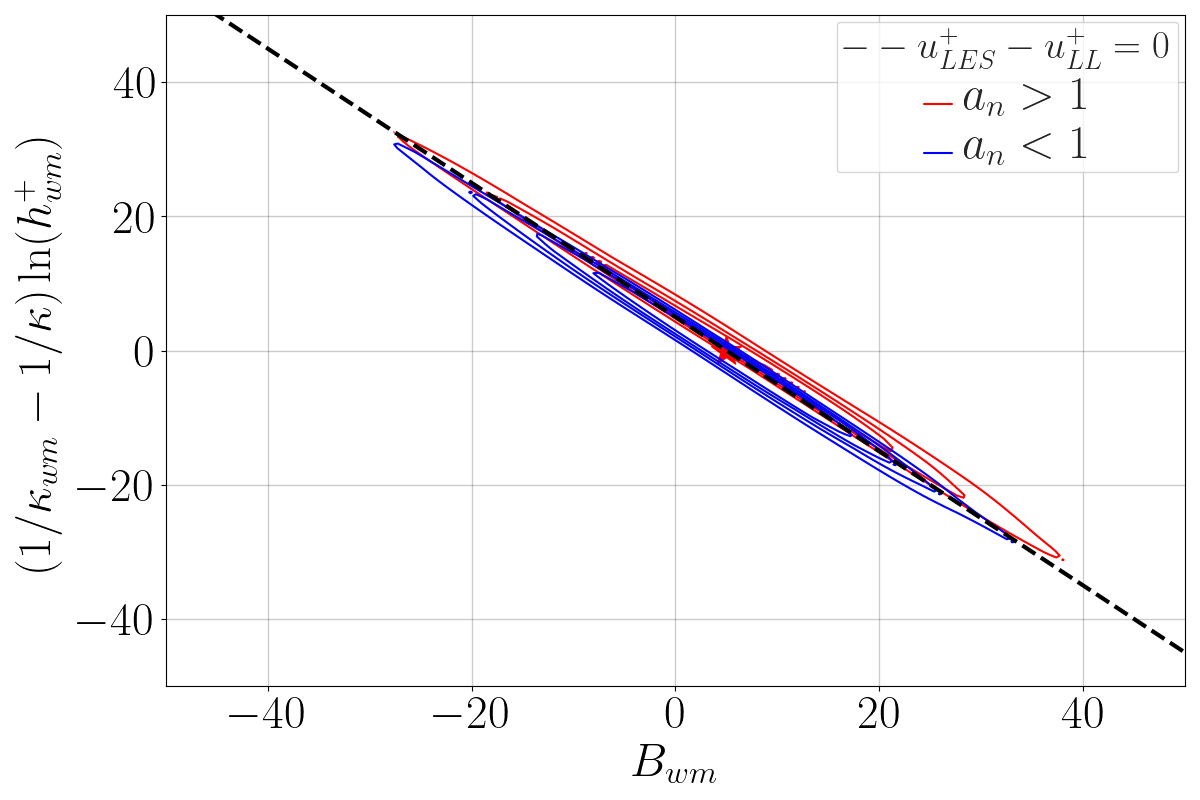}  &  \includegraphics[width=0.4\textwidth]{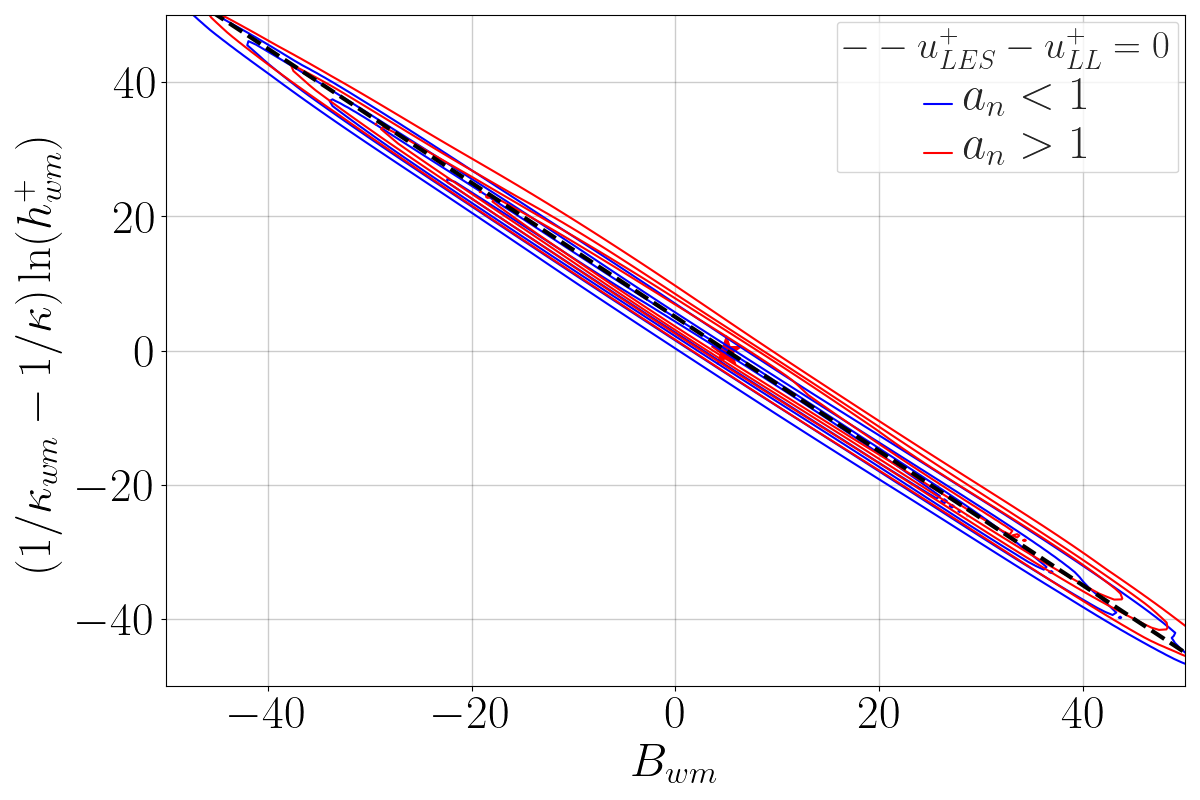}\\
 (c) $Re_\tau=10^{5}$. & (d) $Re_\tau=10^{10}$. \\ 
 \end{tabular}
\caption{\label{fig:analysis2} States-action maps for the VYBA23 WM at (a) $Re_\tau=180$, (b) $Re_\tau=10^3$, (c) $Re_\tau=10^5$ and (d) $Re_\tau=10^{10}$. The following density contours are plotted: $20\%$, $40\%$, $60\%$ and $80\%$.} 
\end{figure}

In this section, we pursue a physical interpretation of the RL WM results, a task that is typically challenging and not frequently undertaken in other ML investigations.
This RL approach is designed in a simple way to control the wall-shear stress and has a limited number of states, allowing for easy visualization of the actions in a 2D plot. 
This is not the case for many other RL models with numerous states, nor for most supervised models that have multiple features, making it challenging to visualize in 2D spaces, and for which, other methods such as the Shapley additive explanations tool (SHAP) need to be used.

Agents learnt a policy through experiences at a Reynolds number of $10^4$. 
They learnt to make decisions regarding the wall-shear stress using only the local values of $\kappa_{wm}$ and $B_{wm}$ to align with the log law. 
If the velocity, $u_{\rm LES}^+$, exceeds the log-law velocity, $u_{\rm LL}^+$, the agents act to increase the wall-shear stress to bring the velocity back to the log-law profile. 
Conversely, if the velocity is below the average, the agents decrease the wall-shear stress. 
The normalization of the first state compensates for an increase in Reynolds number.
The difference between $1/\kappa_{wm}$ and $1/\kappa$ is weighted with $\ln(h_{wm}^+)$ to correct the rotation of the neutral line in the states-action map, rather than using raw $1/\kappa_{wm}$. 
The states-space is designed to be replicable for Reynolds numbers outside the training range, with the exception of when $h_{wm}^+$ is located in the buffer or viscous layers, in which case the VYBA23 WM overpredicts the velocity.
Figure \ref{fig:analysis2} illustrates the agents' policy at Reynolds numbers $180$, $10^3$, $10^5$, and $10^{10}$.
The cluster of points in the states-action map spreads along the neutral line as the Reynolds number increases, due to the increase of $h_{wm}^+$. 
Despite the wider spread of the cluster of points when compared to the learnt case ($Re_\tau=10^4$), the policy of the VYBA23 WM remains consistent for higher Reynolds numbers.

Additionally, the performance of the agents' policies can be evaluated using quantitative measures such as precision and recall, defined as follows:

\begin{equation}
\begin{split}
    \text{precision}(a_n>1) & =\frac{\text{true positive}}{\text{true positive}+\text{false positive}}= \frac{\text{instances of A}}{\text{instances of A}+\text{instances of C}},\\
    \text{recall}(a_n>1) & =\frac{\text{true positive}}{\text{true positive}+\text{false negative}}= \frac{\text{instances of A}}{\text{instances of A}+\text{instances of B}},\\
    %\text{F1 score} &= \sqrt{\text{precision}\times \text{recall}}.
\end{split}
\end{equation}
Similarly, precision and recall can also be defined for action $a_n<1$.
Accuracy can be calculated as:

\begin{equation}
    \text{Accuracy}=\frac{\text{instances of A and D}}{\text{all instances}}.
\end{equation}
The results of these measures are presented in Table \ref{tab:Recall} at Reynolds numbers $180$, $10^3$, $10^5$, and $10^{10}$. 
The VYBA23 WM consistently demonstrates higher precision, recall, and accuracy levels compared to the BK22 WM \cite{bae2022scientific}. 
Although there is a slight decrease in performance at Reynolds number of $10^{10}$, the results are still largely satisfactory.

\begin{table*}
\begin{ruledtabular}
\centering
\caption{\label{tab:Recall} Precision, recall and accuracy scores for BK22 and VYBA23 WMs.}
\begin{tabular}{ccccccc}

  & &\multicolumn{2}{c}{Precision} & \multicolumn{2}{c}{Recall} &  \\
WM&$Re_\tau$ & $a_n>1$ & $a_n<1$ & $a_n>1$ & $a_n<1$ & Accuracy  \\
\hline
\\
\hline
BK22&$180$ & 0.61 & 0.38 & 0.48 & 0.51 & 0.49  \\
\hline
VYBA23&$180$ & 0.91 & 0.92 & 0.90 & 0.93 & 0.92  \\
\hline
\\
\hline
BK22&$10^3$ & 0.61 & 0.40 & 0.49 & 0.52 & 0.50  \\
\hline
VYBA23&$10^3$& 0.92 & 0.91 & 0.90 & 0.93 & 0.91  \\
\hline
\\
\hline
BK22&$10^5$ & 0.66 & 0.40 & 0.51 & 0.55 & 0.53 \\
\hline
VYBA23&$10^5$ & 0.85 & 0.80 & 0.80 & 0.85 & 0.83  \\
\hline
\\
\hline
BK22&$10^{10}$ & 0.57 & 0.42 & 0.45 & 0.53 & 0.49 \\
\hline
VYBA23&$10^{10}$ & 0.64 & 0.65 & 0.67 & 0.62 & 0.64  \\

\end{tabular}
\end{ruledtabular}
\end{table*}

The agents' action amplitude and time-step have a similar impact on wall-shear stress fluctuations (see Figure \ref{fig:taurms}). 
To find a similarity between these two parameters, we derived an equation showing that the ratio of action amplitude to time-step must be the same for both WMs. 

Considering two RLWMs 1 $(\Delta a=a_1;\Delta t =b_1dt)$ and 2 $(\Delta a=a_2;\Delta t =b_2dt)$, between $t$ to $t+ \delta t$, the agents act on the shear-stress as: 

\begin{equation}
    \tau_{t+\delta t} = (1 + a_1)^{\delta t/(b_1 dt)} \tau_t = (1 + a_2)^{\delta t/(b_2 dt)} \tau_t, 
\end{equation}
so that:

\begin{equation}
    \frac{1}{b_1} \ln{(1 + a_1)} =\frac{1}{b_2} \ln{(1 + a_2)}.
\end{equation}
Considering that $a_1<<1$:

\begin{equation}
    \frac{a_1}{b_1}=\frac{a_2}{b_2}.
\end{equation}

\begin{figure}
\centering
\includegraphics[width=0.4\textwidth]{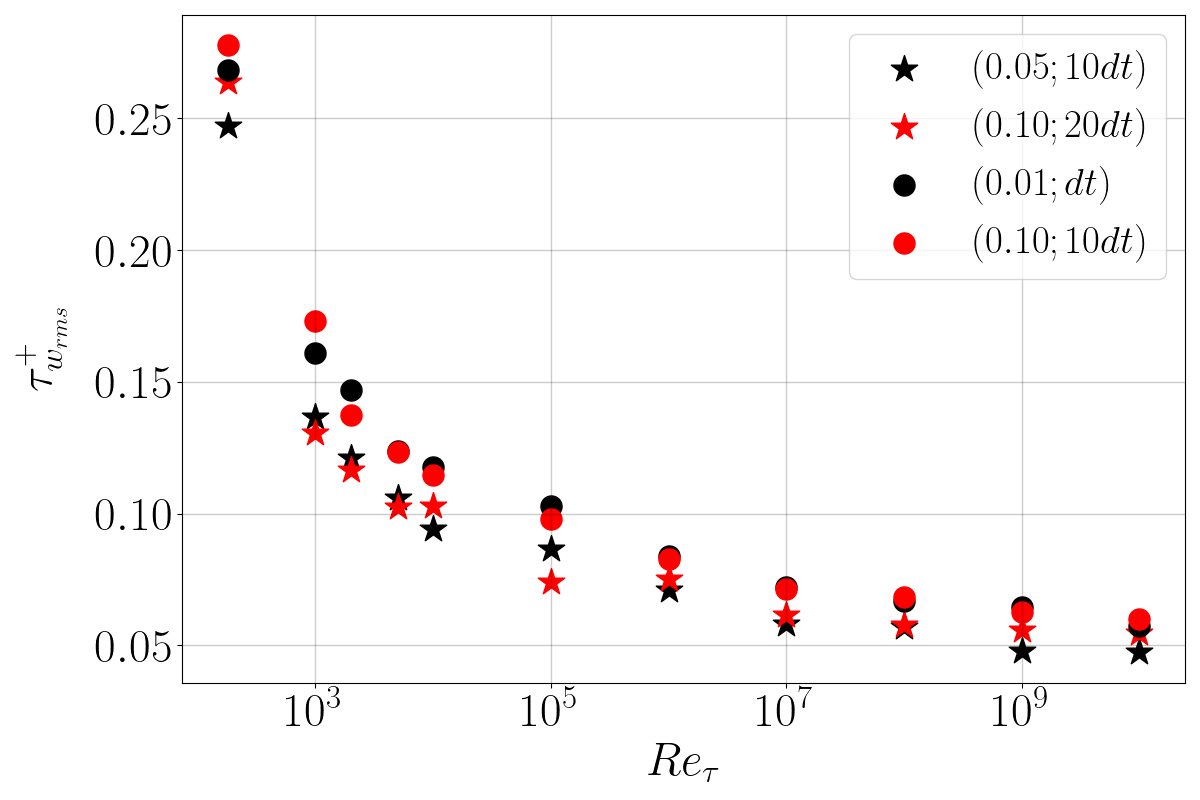}
\caption{ \label{fig:simi_proof} Wall-shear stress fluctuations as a function of friction Reynolds number. We test the similarity rule between two pairs of RLWMs varying agents' parameters: $(0.05;10dt)$ with $(0.10;20 dt)$ and $(0.01;dt)$ with $(0.10;10dt)$. The number of agents is set to $N_{\rm agents}=256$.  } 
\end{figure}

We tested this rule for two cases of action amplitude ratios ($a_1/a_2=2$ and $a_1/a_2=10$).
Results are quite conclusive since curves overlap in both cases (see Figure \ref{fig:simi_proof}).
However, it is more computationally efficient to increase the action range rather than the time-step to have consistent effects. 

\begin{figure}
\centering
\begin{tabular}{ccc}
  \includegraphics[width=0.32\textwidth]{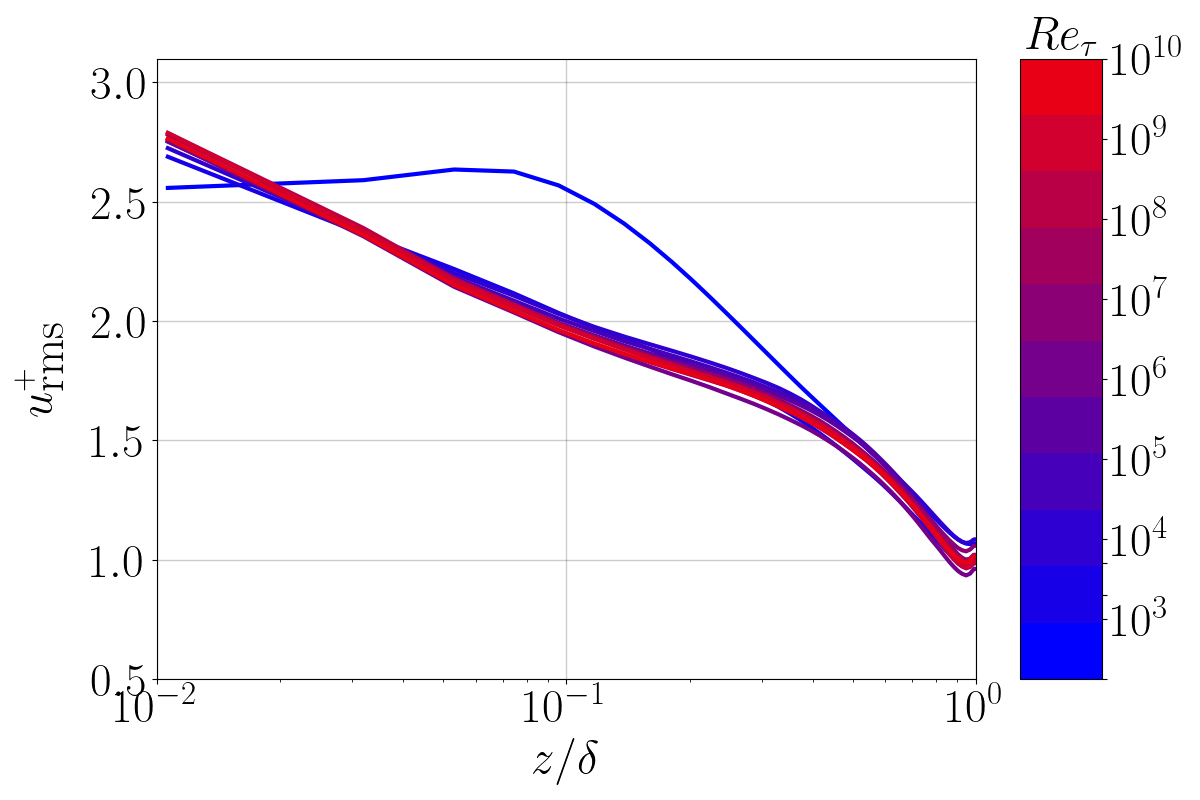}   &  \includegraphics[width=0.32\textwidth]{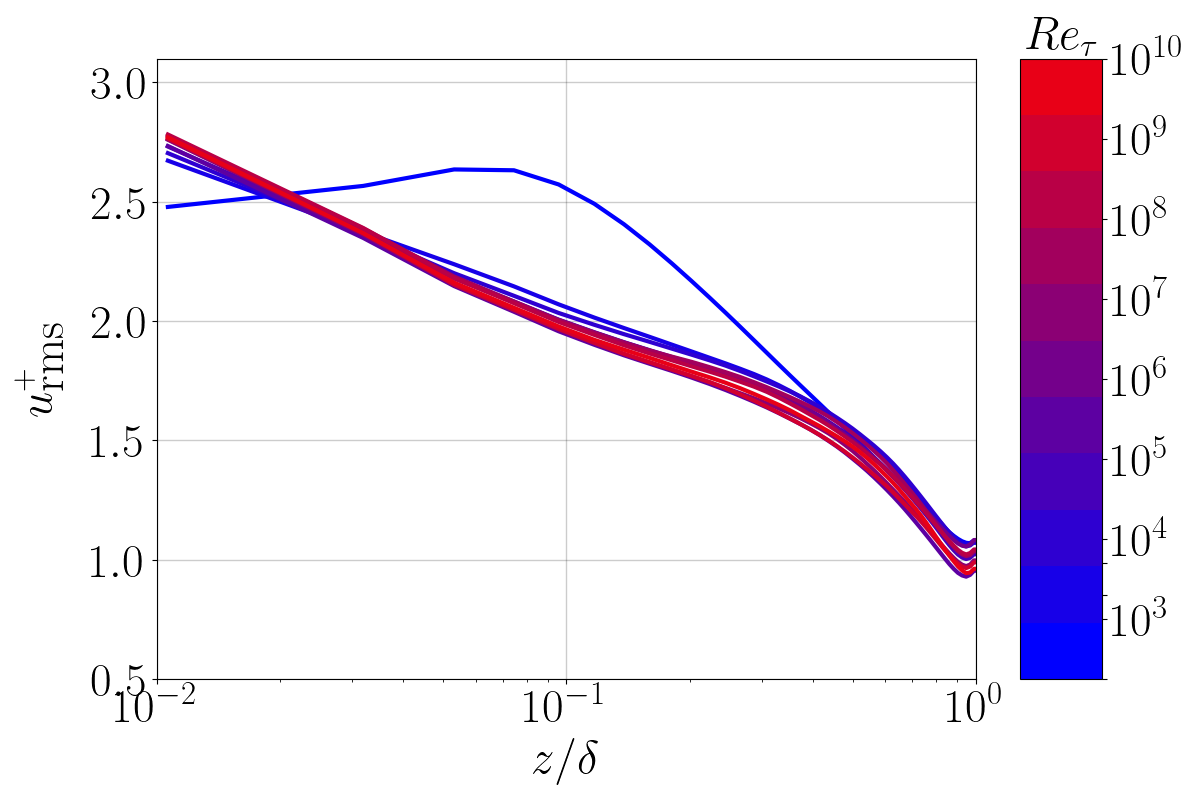}&
  \includegraphics[width=0.32\textwidth]{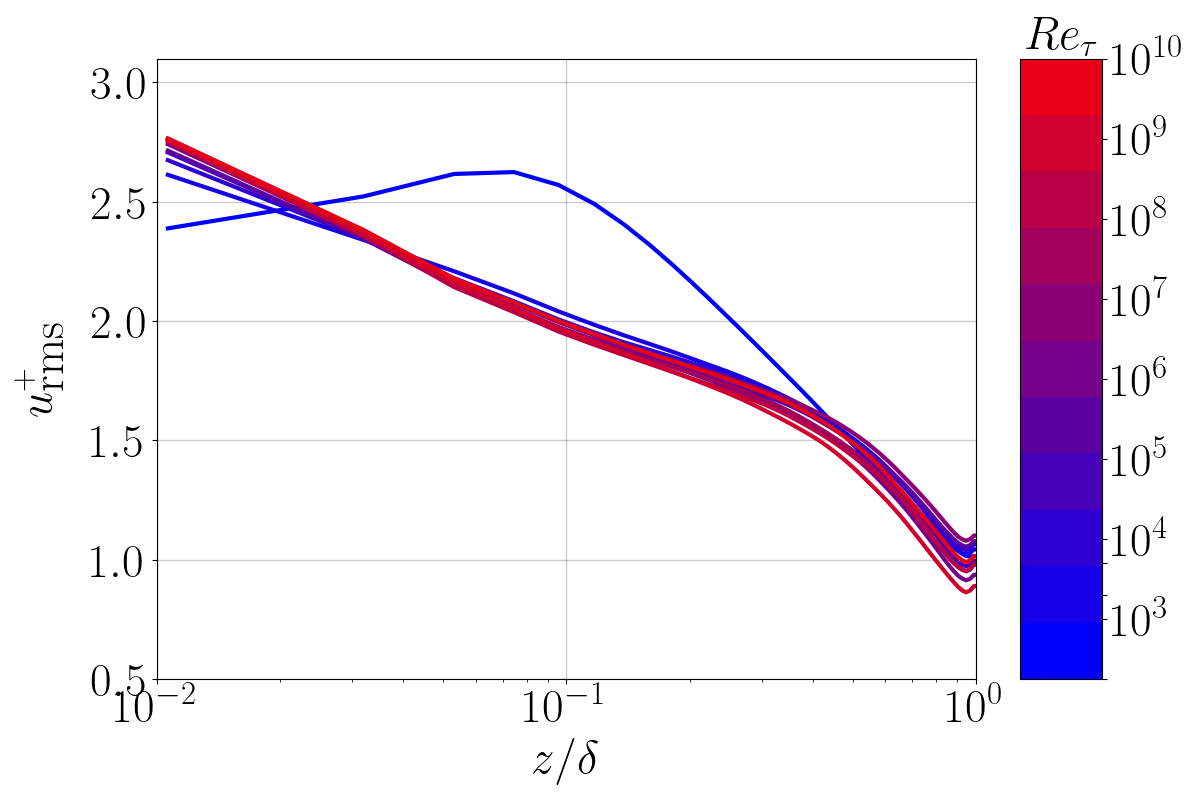}\\
  (a) $\Delta a = 0.025$.  & (b) $\Delta a = 0.05$. & (c) $\Delta a = 0.10$.
\end{tabular}
\caption{ \label{fig:urms_da} Velocity fluctuations for (a) $\Delta a = 0.025$, (b) $\Delta a = 0.05$, (c) $\Delta a = 0.10$. Others parameters are set to the baseline configuration ($\Delta t= 10 dt$ and $N_{\rm agents}=256$, see Table \ref{tab:detail})} 
\end{figure}

\begin{figure}
\centering
\begin{tabular}{ccc}
  \includegraphics[width=0.32\textwidth]{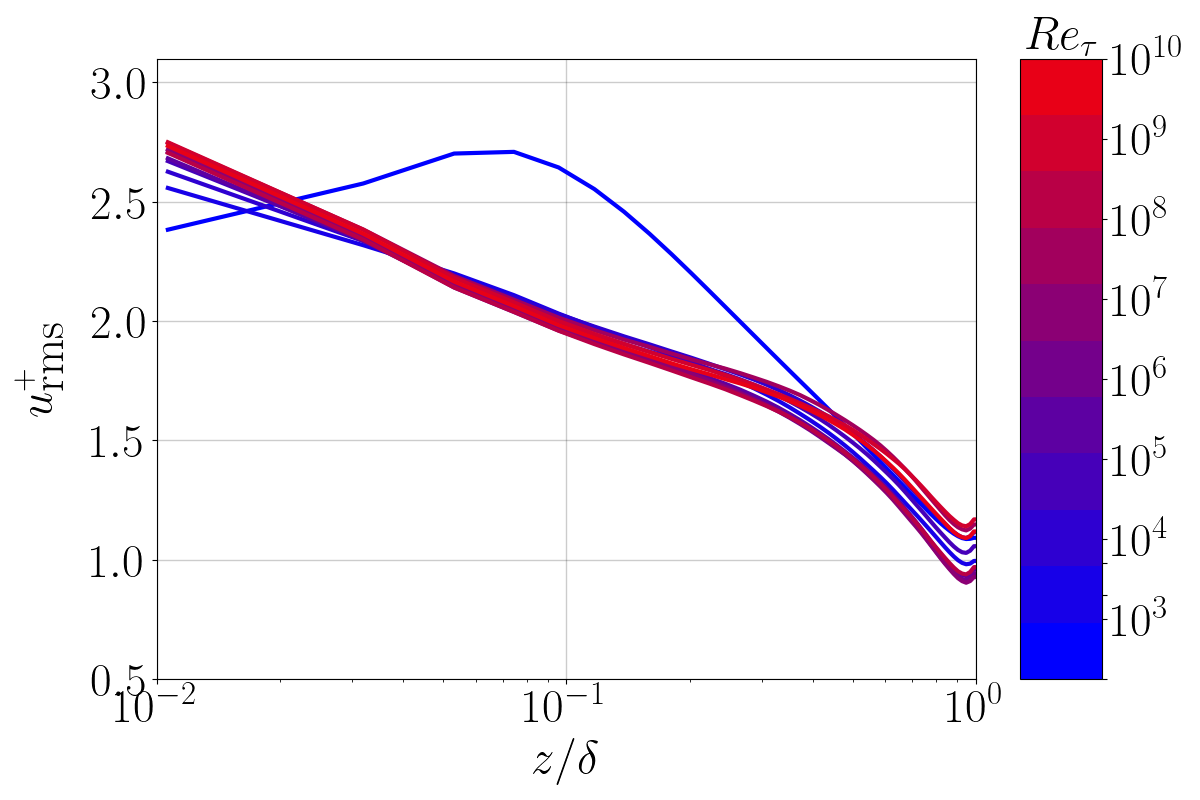}   &  \includegraphics[width=0.32\textwidth]{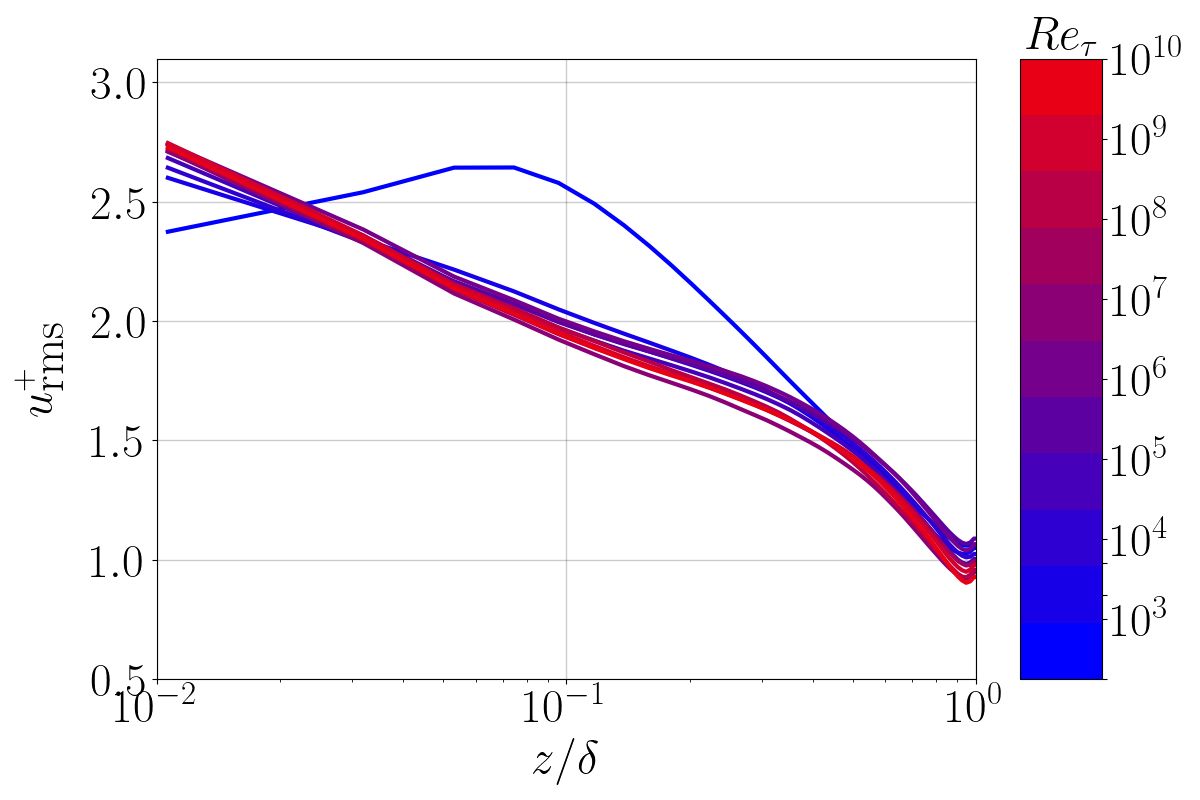}&
  \includegraphics[width=0.32\textwidth]{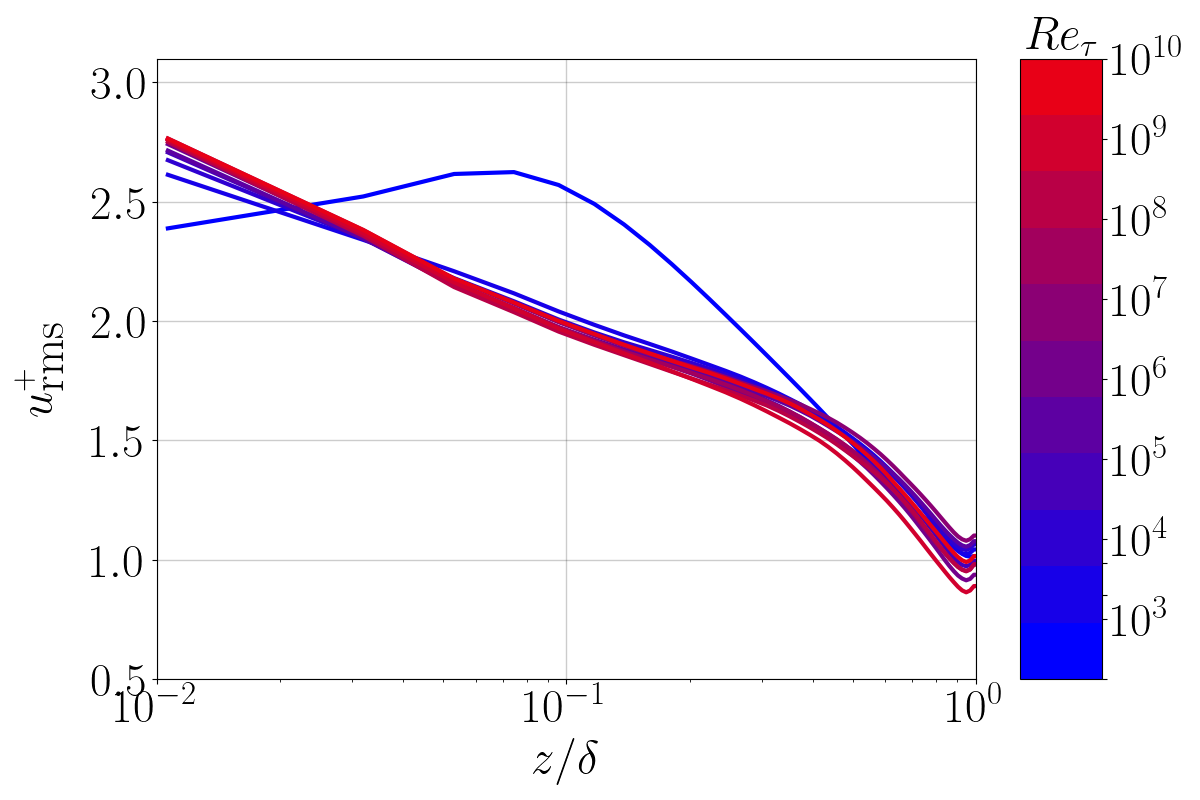}\\
 (a)  $\Delta t = dt$.  &  (b) $\Delta t = 5dt$. & (c) $\Delta t = 10dt$.
\end{tabular}
\caption{ \label{fig:urms_dt} Velocity fluctuations for (a) $\Delta t = dt$, (b) $\Delta t = 5dt$, (c) $\Delta t = 10dt$. Others parameters are set to the baseline configuration ($\Delta a = 0.10$ and $N_{\rm agents}=256$, see Table \ref{tab:detail})} 
\end{figure}

\begin{figure}
\centering
\begin{tabular}{ccc}
  \includegraphics[width=0.32\textwidth]{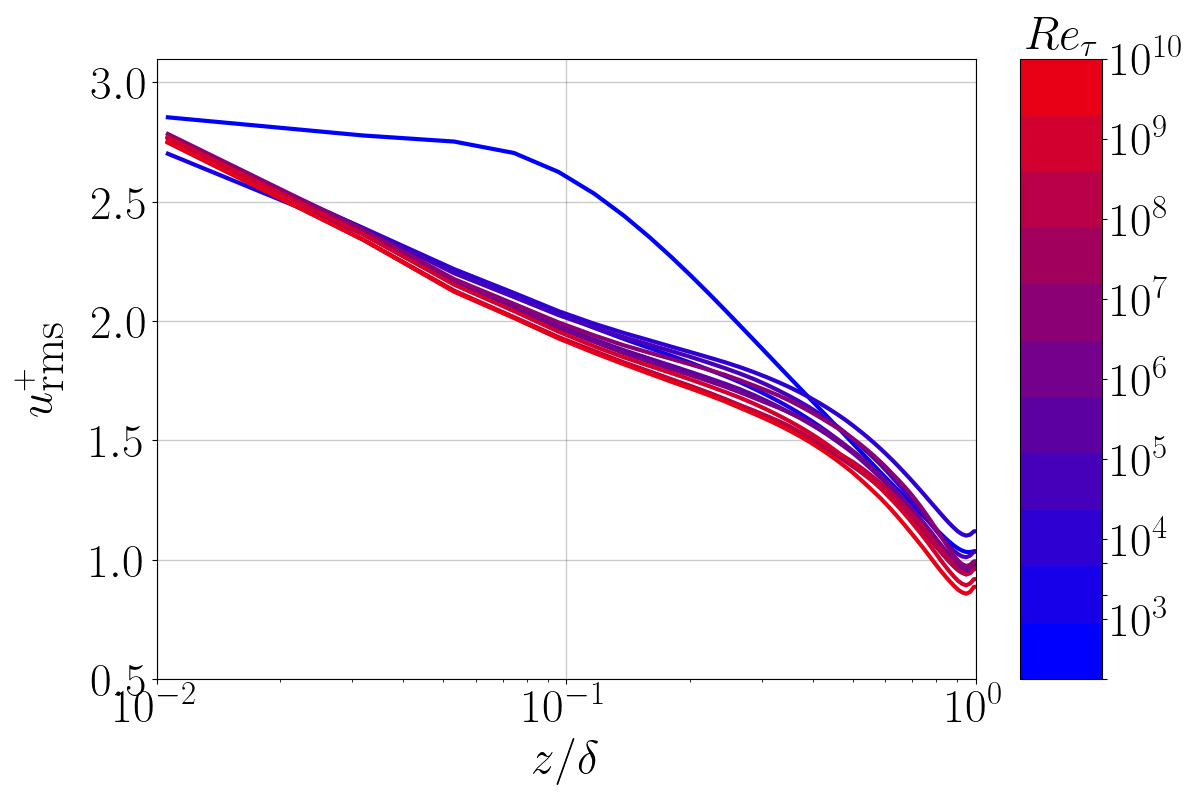}   &  \includegraphics[width=0.32\textwidth]{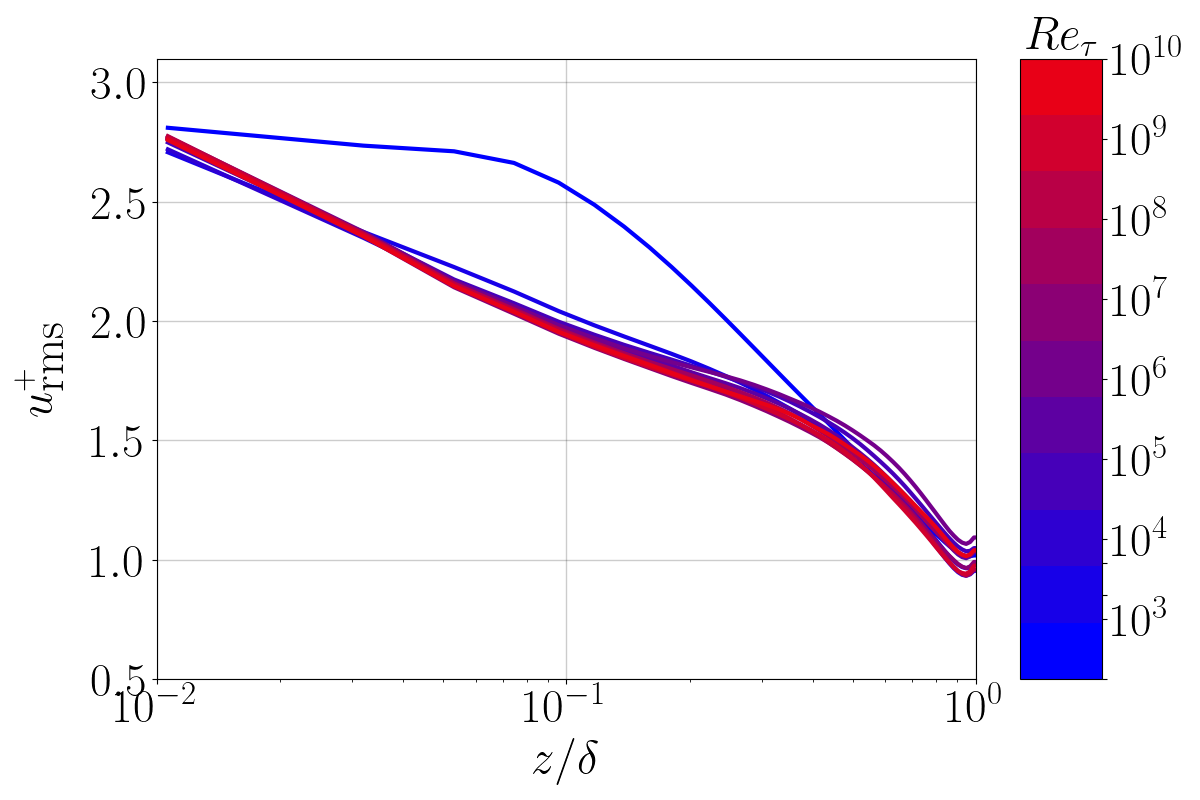} & \includegraphics[width=0.32\textwidth]{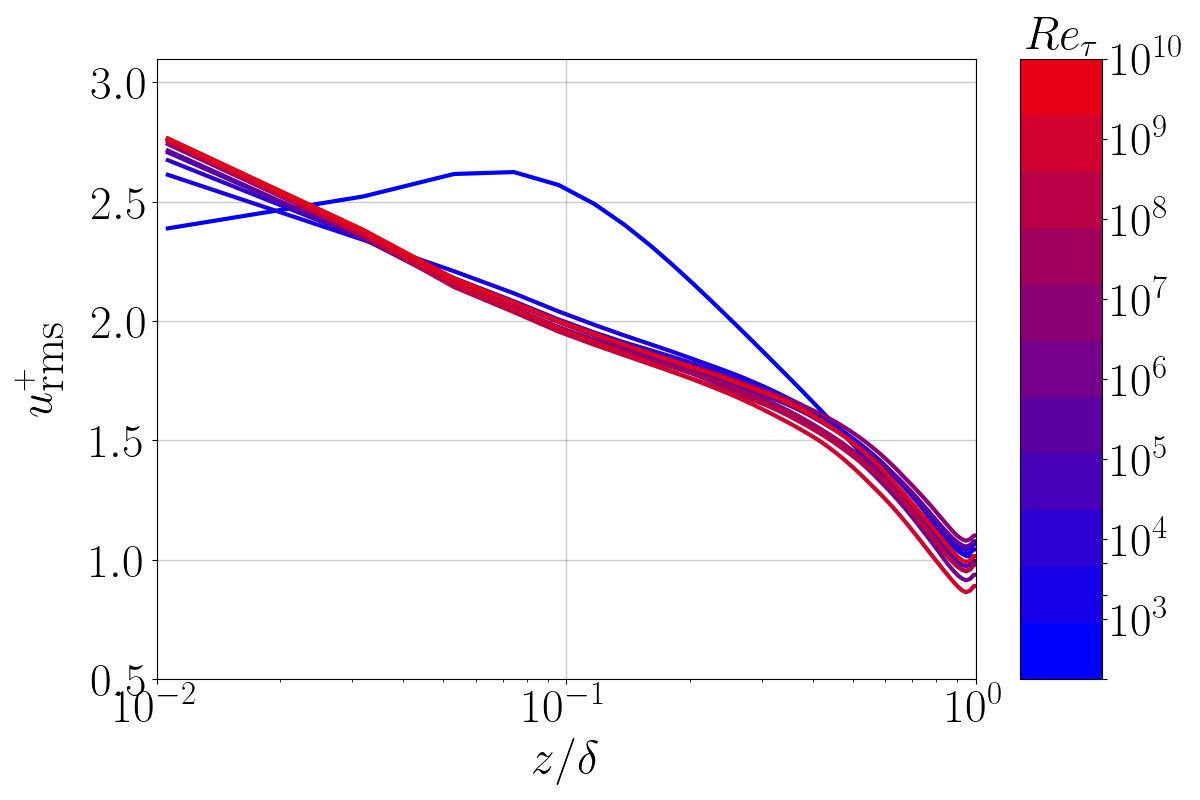}\\
   (a) $N_{\rm agents} = 64$.  & (b) $N_{\rm agents} = 128$. & (c) $N_{\rm agents} = 256$.
\end{tabular}
\caption{ \label{fig:urms_Nagents} Velocity fluctuations for (a) $N_{\rm agents} = 64$, (b) $N_{\rm agents} = 128$, (c) $N_{\rm agents} = 256$. Others parameters are set to the baseline configuration ($\Delta a = 0.10$ and $\Delta t= 10 dt$, see Table \ref{tab:detail})} 
\end{figure}

It is crucial to be mindful of the consequences of changing both the action range and time-step on the flow field, as these changes can impact the Reynolds-number similarity, as shown in Figures \ref{fig:urms_da}, \ref{fig:urms_dt}, and \ref{fig:urms_Nagents}. 
Although the curves follow a very similar trend and almost collapse, it is noted that all three parameters have an effect on it. 
A reduction in the action range leads to more similar results, while for the number of agents or the time-step, there is likely an optimal value somewhere in the middle of the tested range.
Based on these results, it is recommended to reduce the action range, or to set the time-step to $\Delta t = 5dt$, which results in wall-shear stress fluctuations that are closer to the filtered DNS (Figure \ref{fig:taurms}), less log-layer mismatch (Figure \ref{fig:mismatch_LL}) and less impact on the Reynolds number-independence of velocity fluctuations (Figure \ref{fig:urms_dt}). 
However, the relationship between agent's parameters and Reynolds-number similarity is not yet well understood, and this decision may largely depend on the numerical solver used.

\section{\label{sec:conclusion} Concluding remarks}

This paper presents a new WM (VYBA23) based on RL that use agents to control the  wall-shear stress in a flow. 
The model has been developed after an analysis of a previous RLWM (BK22) that was found to produce a log-layer mismatch at large Reynolds numbers due to their choice of states \cite{vadrot2022survey}. 
The VYBA23 model outperforms the BK22 WM by normalizing one of the states and is able to accurately predict the log law for large Reynolds numbers (up to $Re_\tau = 10^{10}$). 
It has also been tested for its ability to predict wall-shear stress fluctuations and has been found to be consistent with filtered DNS \cite{yang2017log} and with the EWM \cite{kawai2012wall}.

The effect of changing the parameters of the agents, such as the action range ($\Delta a$), time-step ($\Delta t$), and the number of agents ($N_{\rm agents}$), was also studied. 
The predictions of the log law remain unchanged, but it was found that these parameters can affect wall-shear stress fluctuations and velocity fluctuations. 

This research provides promising results as the VYBA23 model was trained using only a single Reynolds number ($Re_\tau= 10^4$) without the need for high-fidelity data, yet it still demonstrates good performance in a much broader range.
RL algorithms surpass supervised ML as they work directly with a reward rather than output error, eliminating the need for filtered DNS data and the inconsistency problem between the \textit{a priori} and \textit{a posteriori} computations of the inputs of the model.

Physical knowledge can be incorporated into these models through the use of rewards and states, as proven by the works of Ref. \citenum{bae2022scientific} and this research. 
By shaping the states, the way the RL algorithm represents the environment can be controlled and the dependence on the Reynolds number can be reduced. 
Furthermore, the states-action map, which serves as the policy of the agents, provides a clear understanding of the actions taken by the RL model. 
This eliminates the black-box nature of traditional ML models, providing a more interpretable solution.

To implement this type of ML model, a coupling of the RL library with the CFD solver is required, which can be challenging due to the need for communication between different programming languages. 
In the future, if the superiority of these models is established, it may be worthwhile to explore more efficient methods for extracting the trained model, thus eliminating the need for coupling with the RL library.
Such a development could increase exchange and testing of RLWMs within the community, thereby improving their adoption and confidence in their use.

\blue 
The strength of the RL approach was discussed in detailed in Ref.  \citenum{bae2022scientific}.
Conventional empirical models, RLWMs, and other ML WMs all have their own strengths and weaknesses. 
The purpose of this paper is not to establish the superiority of the RL approach over other approaches.
Rather, we aim to address an issue identified in Ref. \citenum{bae2022scientific}, namely, the log-law recovery.
We consider the recovery of the log law to be an essential step towards creating more general WMs---if a model cannot capture the log law, it would be hard for it to capture other flow phenomena.
\black
RL has demonstrated its potential in solving very large dimensional problems, with the ability to handle up to $10^{170}$ states in the game of Go\cite{maddison2014move}, representing a significant achievement in the field of ML. 
It is believed that this capability could be useful in turbulence modeling, which is a high dimensional problem due to the wide range of scales that increases with the Reynolds number. 
Such an approach could prove particularly valuable in predicting complex flow problems, including separated flows, where significant progress could be made.

\begin{acknowledgments}
This research is supported by the Independent Research Fund Denmark (DFF) under the Grant No. 1051-00015B.
Yang acknowledges US Office of Naval Research under contract N000142012315, with Dr. Peter Chang as Technical Monitor. 

\end{acknowledgments}

\section*{Conflict of interest}
The authors have no conflicts to disclose.

\section*{Data availability statement}
The data that support the findings of this study are available from the corresponding author upon reasonable request.

\bibliography{main}

\end{document}